\documentclass{ws-procs9x6}
\usepackage{hyperref}
\usepackage[utf8]{inputenc}
\usepackage{graphicx}
\usepackage{amsmath}
\usepackage{color}
\usepackage{units}
\usepackage{ifthen}

\newcommand{\doix}[2]{\href{http://dx.doi.org/#2}{#1}}%
\newcommand{\arxiv}[2][]{%
  \ifthenelse{\equal{#1}{}}{%
    \href{http://arxiv.org/abs/#2}{\texttt{arXiv:#2}}%
  }{%
    \href{http://arxiv.org/abs/#2}{\texttt{arXiv:#2 [#1]}}%
  }%
}%

\newcommand{\hlvl}[3]{#1#2\textsubscript{#3}}

\begin{document}
\title{\uppercase{The quest for the proton charge radius}}
\author{\uppercase{István Angeli}}
\address{Institute of Experimental Physics, University of Debrecen, Hungary}

\begin{abstract}
A slight anomaly in optical spectra of the hydrogen atom led \emph{Willis E.~Lamb} to the search for the proton size.
As a result, he found the shift of the \hlvl{2}{S}{1/2} level, the first experimental demonstration of quantum
electrodynamics (QED). In return, a modern test of QED yielded a new value of the charge radius of the
proton. This sounds like \emph{Baron Münchausen}’s tale: to pull oneself out from the marsh by seizing his own
hair. An independent method was necessary. Muonic hydrogen spectroscopy came to the aid. However, the
high-precision result significantly differed from the previous – electronic – values: this is (was?) the \emph{proton
radius puzzle} (2010-2020?). This puzzle produced a decade-long activity both in experimental work and
in theory. Even if the puzzle seems to be solved, the \emph{precise} determination of the proton charge radius
requires further efforts in the future.
\end{abstract}

\section{The Dirac equation; anomalies in hydrogen spectra (1928–1938)}\label{sec:Diraceq}
\begin{figure}
\noindent\hfil\includegraphics[scale=.4]{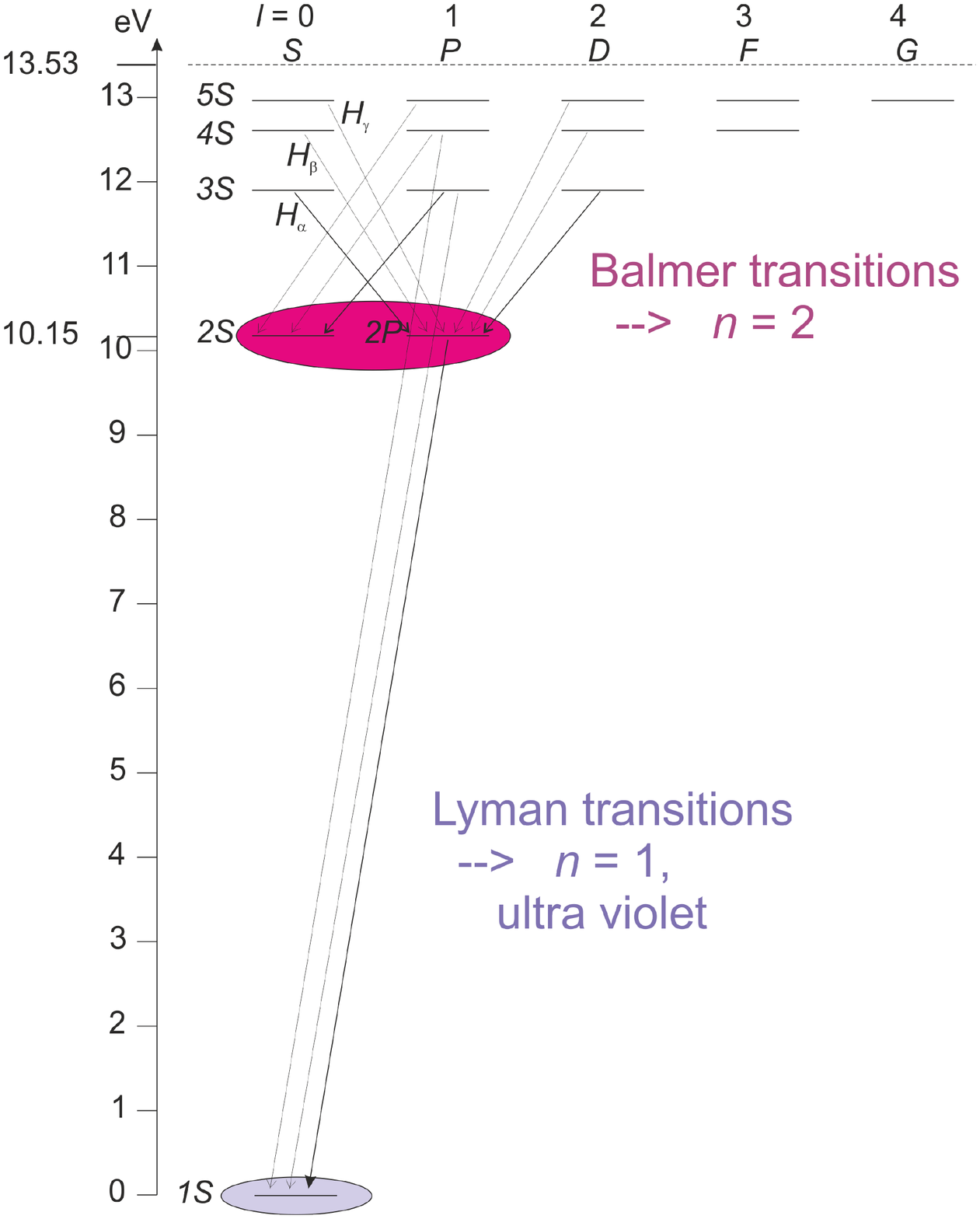}
\caption{Level system of the hydrogen atom in the \emph{Bohr} model.}
\label{fig1}
\end{figure}

\begin{figure}
\noindent\hfil\includegraphics[scale=.4]{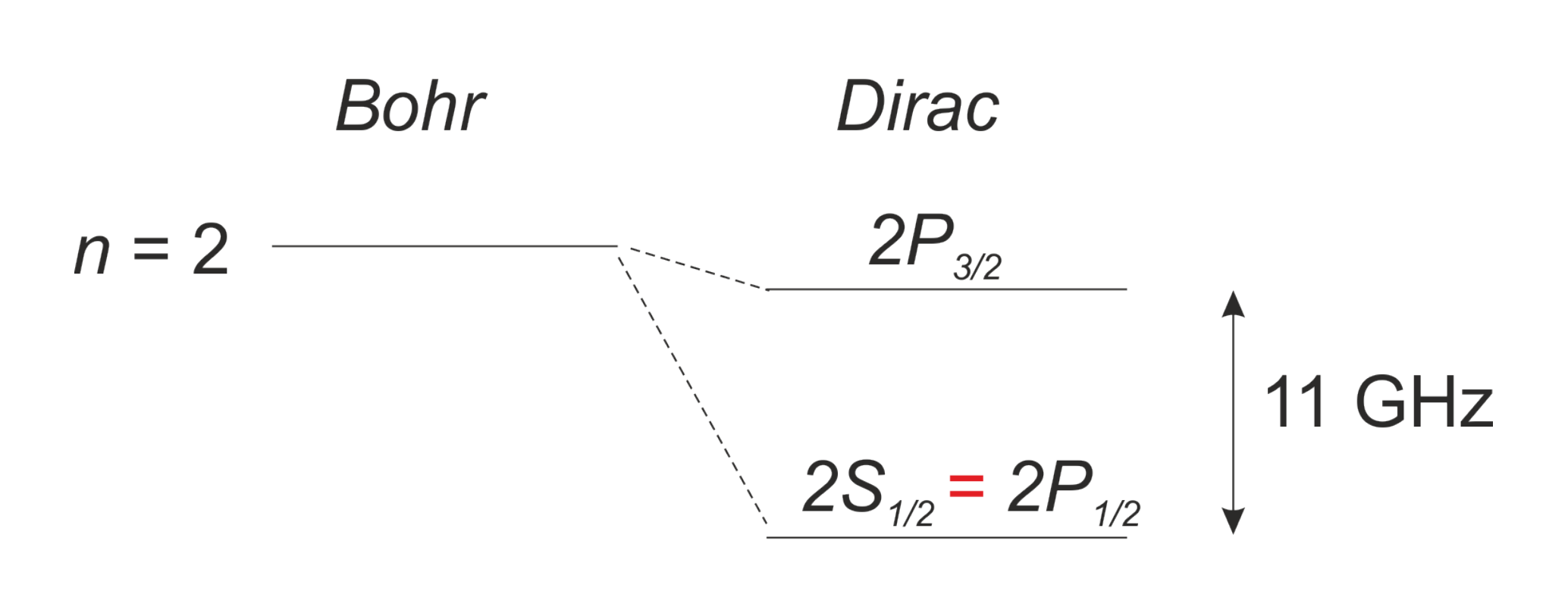}
\caption{Part of the hydrogen atom level system according to \emph{Dirac}; \emph{HFS} omitted.}
\label{fig2}
\end{figure}

In 1928, \emph{P.A.M.~Dirac} published his relativistic wave equation implying two important consequences:
\begin{enumerate}
\item The electron has an intrinsic magnetic dipole moment $\mu_{e} = 1\times\mu_B$ ($\mu_B$: Bohr magneton) in agreement with the experiment (1925: \emph{George Uhlenbeck}, \emph{Samuel Goudsmit}).
\item If in the hydrogen atom the electron moves in the field of a Coulomb potential  $V(r)\sim 1/r$,  then its energy $E(n,j)$ is determined by the principal quantum number $n$ and the total angular momentum quantum number $j$, but not by the orbital angular momentum $l$ and spin $s$, separately.
\end{enumerate}
This means a change in the energy level system of the Bohr model, see Figs\ \ref{fig1}. and \ref{fig2}.

According to \emph{Dirac}’s theory the energies of the states \hlvl{2}{S}{1/2} and \hlvl{2}{P}{1/2}  are equal, while that of \hlvl{2}{P}{3/2} is higher by about  $40\,\mu{\rm eV}$;  this is the fine structure (FS), Fig. \ref{fig2}. This energy difference corresponds to a frequency of  ~11 GHz or a wave length of  $3\,{\rm cm}$.

\begin{figure}
\noindent\hfil\includegraphics[scale=.4]{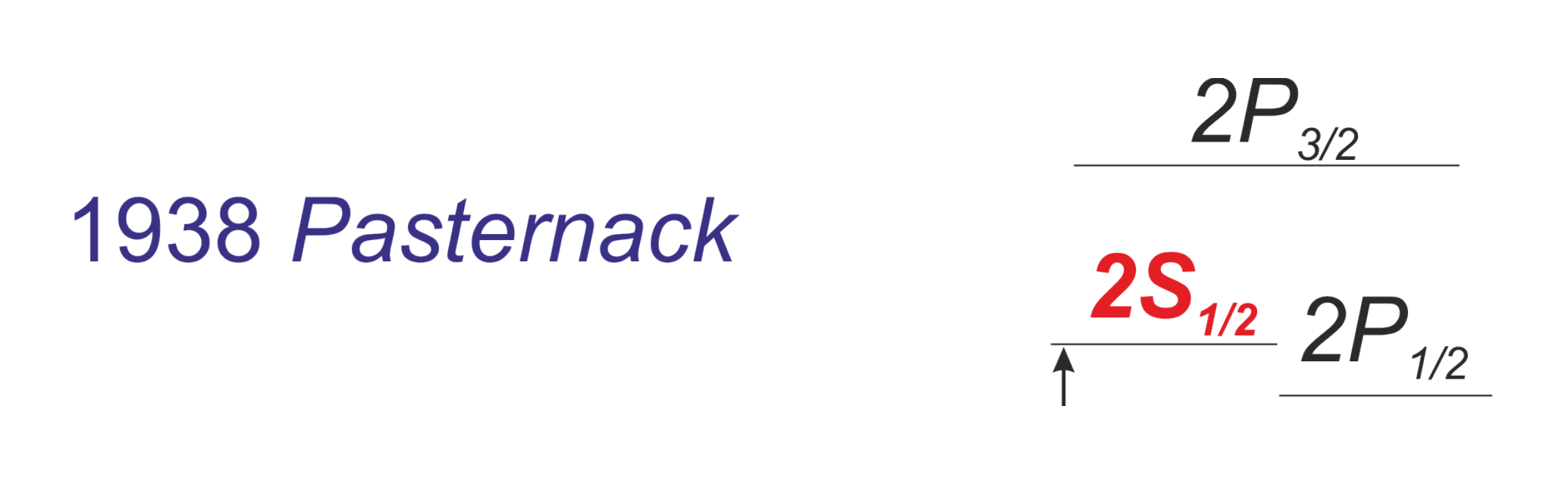}
\caption{\emph{Pasternack}’s idea to explain the Balmer anomaly.}
\label{fig3}
\end{figure}

In those years, the Balmer transitions to the states $n = 2$ were easy to measure; this is the red part of the visible region. In the years ’30, the measurements approved the prediction of the relativistic theory; in some cases, however, there were small deviations, but just at the limit of significance. In 1938, \emph{Simon Pasternack} has shown that these small anomalies can be interpreted assuming that the states \hlvl{2}{S}{1/2} and \hlvl{2}{P}{1/2} do not coincide exactly, but the former is a bit higher\cite{Pa38}, Fig.\ \ref{fig3}.  But he did not concern himself with the origin of the level shift.

\section{The \emph{Lamb} experiment (1938-1947)}\label{sec:LambExp}
In the same year, \emph{Willis E.~Lamb,~Jr.} finished his PhD work under \emph{Robert Oppenheimer}’s supervision.  This work\cite{La38} was closely related to the effect of the ``dynaton'' ($\pi$‑meson) cloud on the structure of nucleons, as suggested by \emph{Hideki Yukawa} in 1935.  (Nota bene: in 1933, \emph{Otto Stern} measured the magnetic dipole moment of the proton to be $2.79\times\mu_N$, i.e., the proton may not be a structureless Dirac particle; Nobel Prize 1943.)  \emph{Lamb} inferred that because of the meson cloud the $V(r) \sim 1/r$ dependence is not valid at small $r$ values. For the \hlvl{2}{S}{1/2}  electrons with $\ell = 0$ orbital angular momentum this results in a weaker binding, i.e., higher energy level compared to \hlvl{2}{P}{1/2}. However, he could not investigate the problem in more detail because of World War II (1939 – 1945).  During the war, some of the physicists -- among them \emph{Lamb} --, had the task of developing the radar detection against aircrafts. \emph{Lamb} especially, investigated the effect of water vapour on the absorption and scattering of microwaves. This practical experience proved to be very useful later in his research work.

After the war, life returned in its peaceful bed.  In summer 1946, \emph{Lamb} --preparing himself for a summer course--, studied \emph{Gerhard Herzberg}’s classical book on molecular spectroscopy. Here he came across a chapter relating the unsuccessful investigation of the $n = 2$ level of the hydrogen atom. He considered that using up-to-date radar technics he will be able to measure the shift of the \hlvl{2}{S}{1/2} level.  To realize his idea, he persuaded his young colleague \emph{Robert C.~Retherford} for collaboration, and they set up the experimental arrangement.

\begin{figure}
\noindent\hfil\includegraphics[scale=.4]{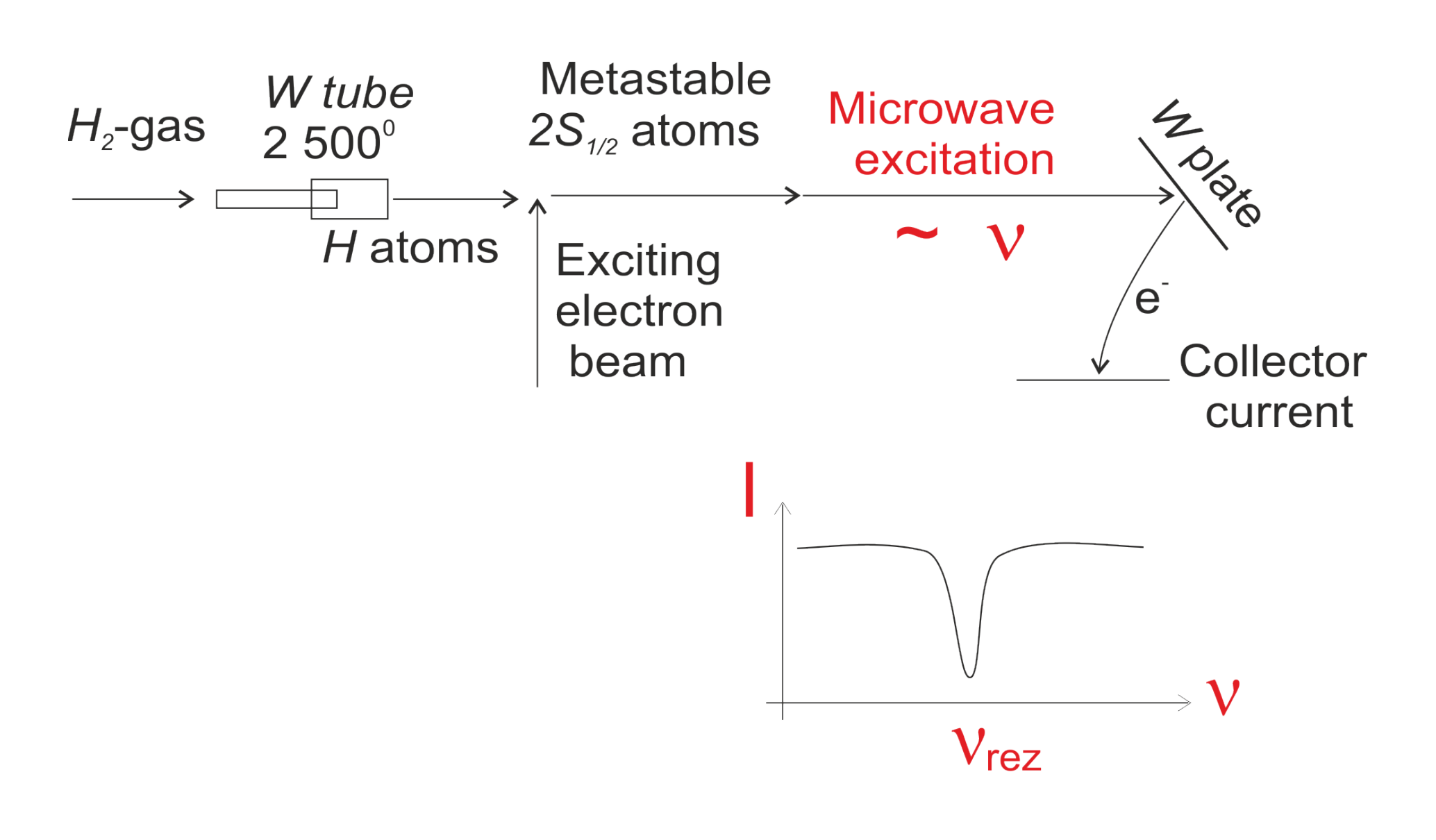}
\caption{Scheme of the experimental arangement.}
\label{fig4}
\end{figure}

The scheme of their setup can be seen in Fig.\ \ref{fig4}.: molecular hydrogen gas passes through the $2500\,{}^\circ{\rm C}$ wolfram tube, here the molecules dissociate to H atoms. These H atoms are excited to the level \hlvl{2}{S}{1/2} by an electron beam. As the probability of radiative transition to the \hlvl{1}{S}{} ground state is very low, \hlvl{2}{S}{1/2} is a \emph{metastable state} with long lifetime. These excited atoms pass through the vacuum tube, at its end they hit a wolfram plate; here the excitation energy is transferred to an electron, which leaves the plate and arrives at the collector. The electron current to this collector is readily measured.

In a section of the arrangement there is a microwave field fed by an oscillator of variable frequency $\nu$.  If this frequency takes the value $\nu_{\rm rez}$ corresponding to the energy difference between \hlvl{2}{S}{} and \hlvl{2}{P}{}, then the transition \hlvl{2}{S}{} $\to$ \hlvl{2}{P}{} takes place, followed by a rapid radiation transition from \hlvl{2}{P}{} to the ground state \hlvl{1}{S}{}.  Then, this H atom does not eject any electron from the wolfram plate: a drop of collector current is observed, Fig.\ \ref{fig4}.\ lower part.

\begin{figure}
\noindent\hfil\includegraphics[scale=.4]{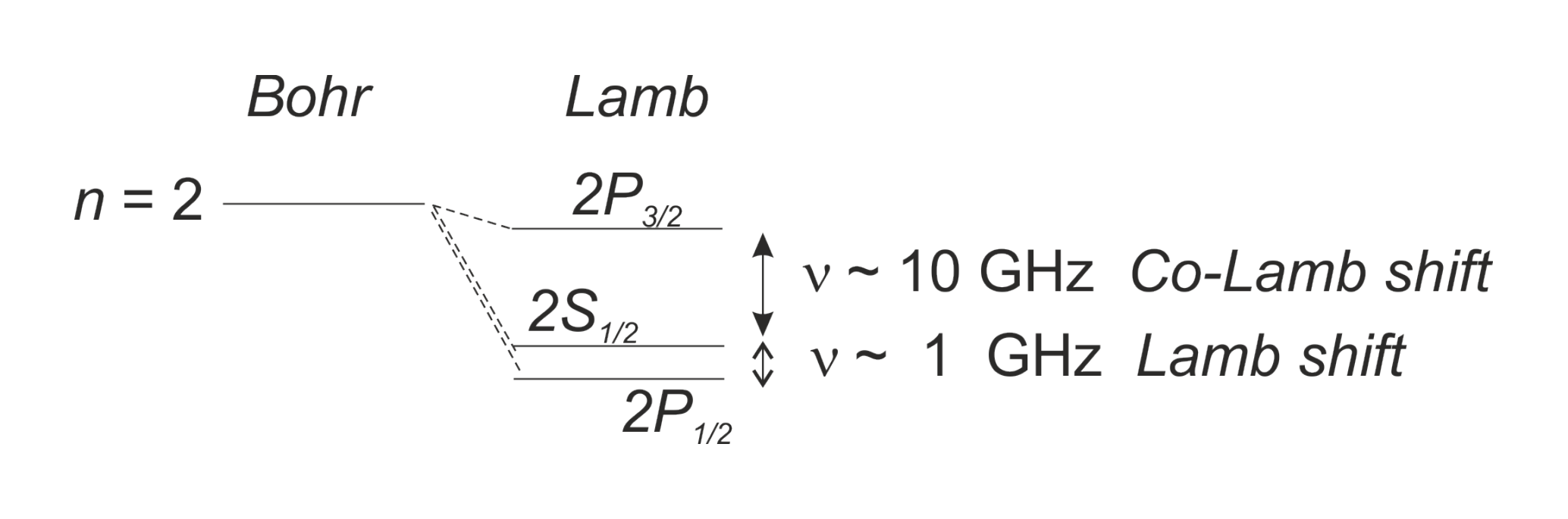}
\caption{The \emph{Lamb shift} and the \emph{co-Lamb shift}.}
\label{fig5}
\end{figure}

In May-June 1947, they got the first results: the state \hlvl{2}{S}{1/2} is only at 10\,GHz from \hlvl{2}{P}{3/2} instead of 11\,GHz\cite{La47}. Thereafter, the energy difference \hlvl{2}{S}{1/2}-\hlvl{2}{P}{1/2} was also measured: $\sim$1\,GHz; this energy difference is called today \emph{Lamb shift}, Fig.\ \ref{fig5}. The \hlvl{2}{S}{1/2}-\hlvl{2}{P}{3/2}  difference is the \emph{co-Lamb shift}\cite{Br93}. These two measurements confirmed \emph{Pasternack}’s suspicion, that the state \hlvl{2}{S}{1/2} is higher than \hlvl{2}{P}{1/2}.  This shift, however, is several orders of magnitude larger than expected from the size of the meson cloud, i.e., from the size of the proton.  This latter is only  $\sim 0.08\,{\rm MHz}$, as calculated later\cite{Sl49}.

\section{Experimental foundations of quantum electrodynamics (1947-1953)}\label{sec:expQED}
\begin{figure}[hb!]
\noindent\hfil\includegraphics[scale=.4]{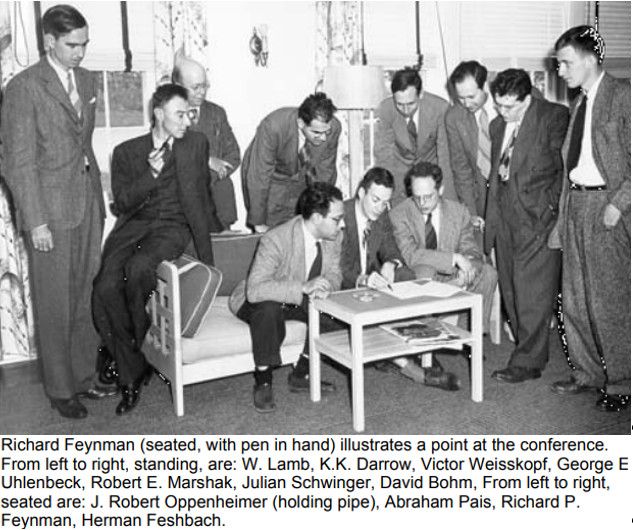}
\caption{Participants of the Shelter Island Conference. (Photo from Ref.\ \citenum{NA07})}
\label{fig6}
\end{figure}

\emph{Lamb} was lucky to participate in a historical event: during 2--4 June 1947, a conference was organized on the actual problems of post-war theoretical physics\cite{Sm96}. This conference took place in Shelter Island near New York. Some of the participants are shown on the photo, Fig.\ \ref{fig6}.

\emph{Shelter Island provided the initial stimulus for the post-war developments in quantum field theory: effective, relativistically invariant computational methods, Feynman diagrams, renormalization theory.  Many believe that the symphony which we later came to call quantum electrodynamics (QED) started with the tuning-up and improvisations heard at a quaint little country inn near the Hamptons in the spring of 1947.}\cite{Sm96}.

As an outcome of the discussions it became clear that the shift of the level \hlvl{2}{S}{} is caused by the interaction between the electron and the quantized electromagnetic radiation field.  On the way home from the conference, \emph{Hans Bethe}\cite{Be47}  performed a simple, non-relativistic calculation on the shift of hydrogen \emph{Dirac}-levels, and he got the result $\Delta(\text{\hlvl{2}{S}{1/2}}) = 1040\,{\rm MHz}$ 
 in very good agreement with the experimental values measured by \emph{Lamb} and \emph{Retherford}.  Was the calculation that easy?  For \emph{Bethe}, it was:

\begin{quote}
\emph{Hans at age four sitting on the stoop of their house, a piece of chalk in each hand, taking square roots of numbers. Bethe learned to reason qualitatively, to obtain insights from back-on-envelope calculations. The key to Bethe's success was in his interpretation of the infinities that arise in the calculation}\cite{Ma20}.
\end{quote}

Owing to the discussions at the Conference, \emph{Lamb} realized the extreme importance of their experiment. After the immediate publication of the first results on one and a half pages\cite{La47}, he performed an improved, painstaking series of experiments, \emph{he changed to sixth gear}, in order to exclude any doubt whatever about the correctness of the experimental results now founding QED.  This work has been published in a series of six dedicated papers in the Physical Review from 1950 to 1953.  A description of readable size has also been published\cite{La51} which can be recommended for the interested physicist reader.

In the same year, \emph{I.I.~Rabi} and his group at Columbia University, N.Y., measured the \emph{hyperfine structure} (\emph{HFS}) of the hydrogen atom\cite{Na47}. The result of the experiment yielded a deviation of  $5\sigma$ from the value calculated theoretically by \emph{Enrico Fermi}.  Because of the simple structure of the hydrogen atom, there was no reason to question the validity of the simple theoretical formula.

\emph{Rabi} sent the peculiar result to \emph{G.~Breit}, who declared that the only consistent removal of the dilemma is that the the electron has an intrinsic magnetic moment of $\mu_e = 1.001\times\mu_B$\cite{Br47}, i.e., higher than the \emph{Dirac} value. \emph{Breit} himself suggested this solution reluctantly, but in his paper he can not put a better argument against it: \emph{-Aesthaetic objections can be raised against such a view}. We understand this reasoning; an integer value were much more \emph{beautiful}:
\begin{quote}
\emph{``Beauty is truth, truth is beauty, -- that is all\\
Ye know on earth, and all ye need to know''}\\
\hspace{4cm}(John Keats: \emph{Ode on a Grecian Urn})
\end{quote}

One had to recourse to experiment again. In order to eliminate the effect of some disturbing interaction between the proton and the electron \emph{P.~Kusch} and \emph{H.M.~Foley} measured the magnetic dipole moment of the peripheral electron in Na, Ga and In atoms\cite{Ku48}. The result of their experiment was  $\mu_e = 1.00119(5)\times\mu_B$.  In this same year, \emph{J.~Schwinger} calculated the QED contribution to the magnetic moment of the electron, and has shown that it is of the order of  $\sim \alpha/2\pi\approx 0.001$\cite{Sc48}, in agreement with the experiments. 

These experiments were so decisive for the proof of quantum electrodynamics that the 1955 Nobel Prize in Physics was awarded to \emph{Lamb} and \emph{Kusch}. Ten years later \emph{Feynman}, \emph{Schwinger} and \emph{Tomonaga} got the prize for the development of the theory.

Note that in 2020, the fine structure in the $n = 2$ states of \emph{antihydrogen} was measured by the \emph{ALPHA Collaboration}\cite{AL20}, and the resulting value for the fine-structure splitting \hlvl{2}{P}{1/2}-\hlvl{2}{P}{3/2} is consistent with the predictions of quantum electrodynamics to a precision of  2\,\%, while the classic Lamb shift in \emph{antihydrogen} \hlvl{2}{S}{1/2}-\hlvl{2}{P}{1/2} is consistent to 11\,\%. This means a positive test of the \emph{charge-parity-time symmetry}.

\section{Discovery of New Worlds}\label{sec:discovery}

\begin{figure}
\noindent\hfil\includegraphics[scale=.4]{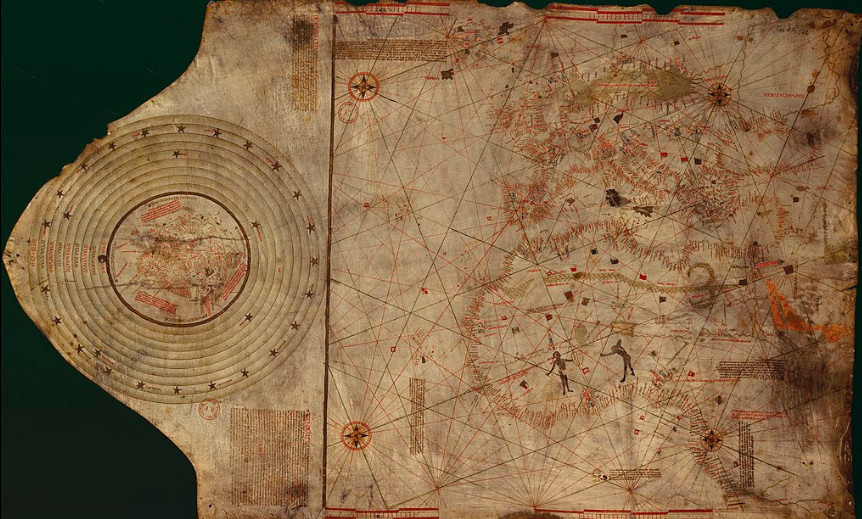}
\caption{This map was drawn circa 1490 in the workshop of \emph{Bartolomeo} and \emph{Christopher Colombo} in Lisboa [Wikipedia]}
\label{figM}
\end{figure}

In 1492, instead of some Marvellous India (Fig.\ \ref{figM}) -- an unheard of continent was discovered.

That's great in science! You have a question mark in your mind's eye. In the search for the answer, you fit together the necessary experimental and theoretical tools, spend a lot of time and work with it--also some bureaucracy--, and \emph{you'll succeed at last}. But sometimes the answer is not what you expected; it may be even more exciting. With \emph{quantum electrodynamics} a new realm opened up. The \emph{Lamb experiment}--originally intended to determine the the proton size--, is not unique. Another example follows.

\subsection{Genesis of modern data analysis}\label{ssec:datanal}
In the last decade of the 18\textsuperscript{th} century, two French astronomers, \emph{Jean Baptiste Joseph Delambre} and \emph{Pierre Francois André Méchain} measured the length of the meridian between Dunkirk and Barcelona to define the meter as one ten-millionth of the distance between the pole and the equator\cite{Al03}. The data of their triangulation measurements were published in the \emph{Base du système métrique décimal} at over two thousand pages by \emph{Delambre}. For centuries, savants had felt entitled to use their intuition and experience to publish their single ``best'' observation as a measure of a phenomenon. 

\emph{Adrien-Marie Legendre} had a special interest in evaluating the heap of data, because originally he was also appointed to the Revolutionary Meridian Project, but he withdraw in favour of \emph{Delambre}. In 1805, \emph{Legendre} suggested a practical solution: the best curve would be the one that minimized the sum of squares of departures of each data point from the curve. The \emph{least-squares method}, the workhorse of modern statistical analysis, also justified choosing the arithmetic mean in the simplest cases. 

Four years later, \emph{Carl Friedrich Gauss} claimed that he had been using the least-squares rule for nearly a decade. They arrived at the method independently: \emph{Legendre} published it first as a practical tool, but \emph{Gauss} showed the method's deeper meaning (predecessor of the \emph{maximum likelihood}?) by showing that it gave the most probable value in those situations where the errors were distributed along a \emph{bell curve} (known today as a \emph{Gaussian} curve).

The probability-based approach prompted \emph{Pierre-Simon Laplace} to show that the method indicated how to distinguish between \emph{random errors} and \emph{constant/systematic errors}.  These years from 1805,  saw  \textbf{the rise of a new scientific theory;  not a theory of nature, but a theory of error}\cite{Al03}{}\textsuperscript{p.\ 307}.

\section{Elastic electron scattering: the Stanford era (1953--1963)}\label{sec:Stanford}
\begin{figure}
\noindent\hfil\includegraphics[scale=.2]{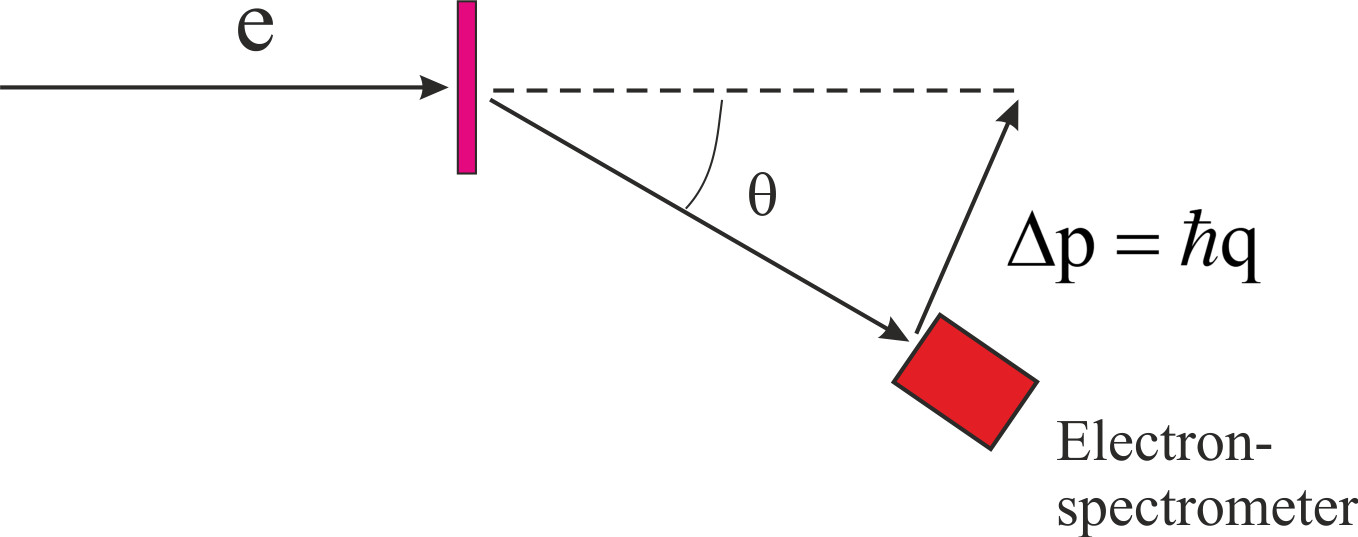}
\caption{Experimental arrangement for fast electron scattering.}
\label{fig6masodik}
\end{figure}

Unquestionably, \emph{Lamb}'s experiments were of great importance for the foundation of quantum electrodynamics. However, the original question, the size of the proton, remained unanswered. In 1953, \emph{Robert Hofstadter} at Stanford University launched a far-reaching experimental program for the determination of charge distributions in atomic nuclei by fast electron scattering\cite{Ho53} (Nobel Prize: 1961). The scheme of the experimental arrangement is shown on Fig.\ \ref{fig6masodik}. Electrons accelerated to several hundred MeV are scattered on nuclei to be investigated; the momentum transfer is $\Delta p=\hbar q$.  The quantity $q$ plays an important role in the analysis of scattering experiments. The scattered electrons are bent vertically to the scattering plane by a $180^\circ$ magnet to reach a Ćerenkov detector. The ways of the 40‑odd tons spectrometer are fastened to a double five-inch anti-aircraft gun mount \emph{``kindly furnished by the Bureau of Ordnance, U.S.\ Navy''}\cite{Ch56}.

\begin{figure}
\noindent\hfil\includegraphics[scale=.3]{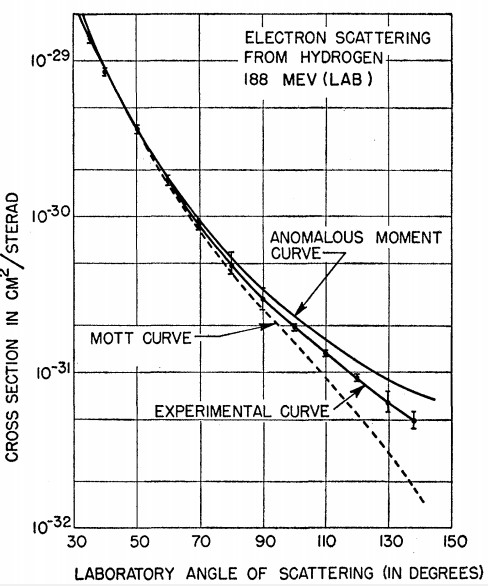} \includegraphics[scale=.15]{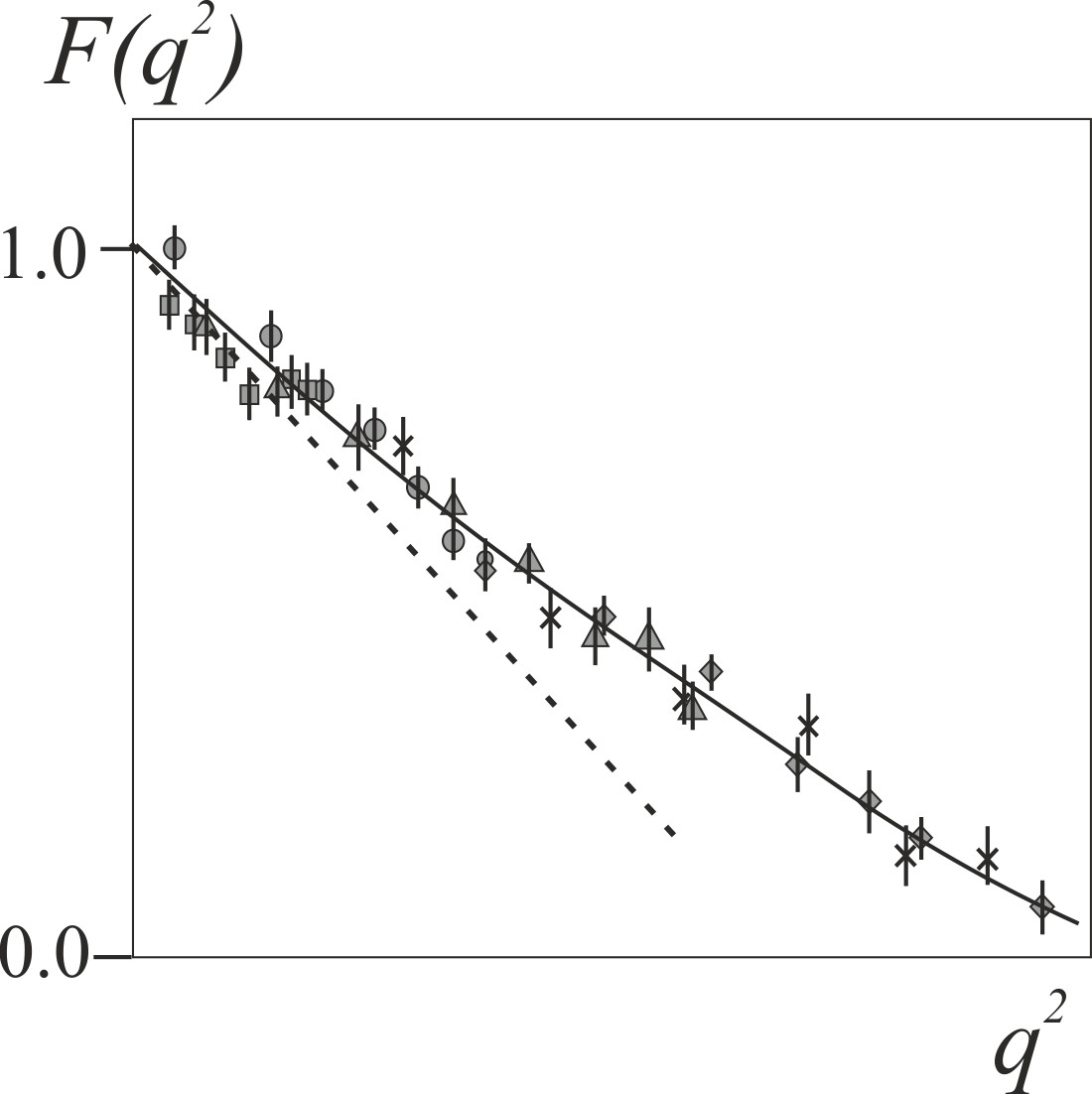} 
\caption{(left) A historic figure, demonstrating the finite size of the proton\cite{Ho55}. (right)~Form~factor $F(q^2)$ of the proton\cite{Ch56}.}
\label{fig7}
\end{figure}

The scattering cross section $\sigma_M(\theta)$ for a point-like nucleus has been calculated by \emph{Nevill Francis Mott}. In 1955, it was observed that the experimental points do not fit either the \emph{Mott} curve or the theoretical curve of \emph{Rosenbluth}, computed for a point charge and a point (anomalous) magnetic moment of the proton, Fig.\ \ref{fig7} (left)\cite{Ho55};  i.e.,
\begin{center}
\textbf{the proton is not point-like!}
\end{center}

In the same paper, the first estimation of the \emph{rms} charge radius is also given\cite{Ho55, Mc56}:
\[
r_p = 0.74(24)\,{\rm fm}\,.
\]

Later, however, the authors note with caution:  \emph{-An equivalent interpretation of the experiments would ascribe the apparent size to a breakdown of the Coulomb law and the conventional theory of electromagnetism}\cite{Ch56}.

The ratio of the measured $\sigma(\theta)$ cross section to $\sigma_M(\theta)$,
\[
\frac{\sigma(\theta)}{\sigma_M(\theta)} = \left|\int\rho(\vec{r}){\rm e}^{i\vec{q}\cdot\vec{r}}{\rm d}^3 \vec{r}\right|^2 = \left|F(\vec{q})\right|^2\,,
\]
is determined by the charge distribution $\rho(\vec{r})$ of the nucleus through the form factor $F(\vec{q})$.  For charge distributions with spherical symmetry $\rho(r)$, the form factor can be expressed by the simpler formula
\[
F(q) = \int \rho(r) \frac{\sin qr}{qr}4\pi r^2{\rm d}r\,.
\]
For \emph{small values} of $q$ this is approximated by the power series
\[
F(q) = 1-\frac{\langle r^2\rangle}{3!}q^2 +\frac{\langle r^4\rangle}{5!}q^4 -\dots+\dots\,.
\]

In his calculation \emph{Mott} assumed that the spin and the magnetic dipole moment of the scattering nucleus is zero. For even-even nuclei this is exactly true, for medium and heavy nuclei approximately. In the case of light nuclei, however, the effect of the magnetic moment of the nucleus on the electron scattering should also be taken into account. Consequently, from the scattering cross section two form factors are derived, an electric $G_E(q)$ and a magnetic $G_M(q)$. These are functions of $q$ derived from the \emph{four-momentum change}. Using the power series of the electric $G_E(q)$ function, the derivative according to $q^2$ yields the mean square radius:
\[
\langle r^2\rangle_p \approx -6\left[\frac{{\rm d}G(q^2)}{{\rm d}q^2}\right]_{q^2=0}\,,
\]
i.e., the slope at $q=0$ is to be determined. (For more precise definitions see Eq.\ (21) in Refs.\citenum{Pa99, Mi19}.) Fig.\ \ref{fig7}.\ (right) presents the data of an early experiment\cite{Ch56}, and demonstrates that the practical slope determination is not at all simple; it contains sources of systematic errors. The uncertainty of the few experimental data at the smallest $q$ values results in rather high error of the slope; see also Ref.\ \citenum{Si14}.

On the other hand, taking the data within a wide $q$ interval, the trend of the points deviates from linearity. One has to assume a \emph{model} to describe the $G_E(q^2)$ dependence. Then, fitting the parameters of the model to the experimental data, the value of the slope at $q=0$ can be extrapolated. The result will depend on the model assumed. A simple, practical model is the \emph{dipole function} $1/(1 + cq^2)^2$, which is the Fourier transform of the exponential density distribution. Subsequent investigations have shown that the dipole function reproduces the trend of cross sections within about $\pm 10\,\%$, see e.g. Ref.\ \citenum{Be71}.

Later, with the development of experimental techniques (higher energy, better electron spectrometers) and theoretical evaluation methods, more series of measurements were performed. Some of them published cross sections and form factors only, others derived \emph{rms} charge radii, too. These latter are\cite{Ch56, Mc60, Bu61}:
\[
r_p = 0.77(10)\,{\rm fm}
\,,\quad  r_p = 0.71(12)\,{\rm fm}
\,,\quad r_p = 0.75(5)\,{\rm fm}
\,.
\]
The early activity of the Stanford group on the proton and neutron is reviewed in Ref.\ \citenum{Ho58}.

\section{Elastic electron scattering; new scenes, new techniques (1963--1980)}\label{sec:elasticNT}
Presumably inspired by the results in Stanford, several laboratories in the world begun electron scattering measurements. At Orsay, the \emph{low‑$q$} measurements resulted in\cite{Du63,Le62}
\[
  r_p = 0.82(2)\,{\rm fm}\,,\quad\quad\quad
r_p = 0.84(4)\,{\rm fm}\,.
\]
In 1963, measurements on the proton as performed in Stanford, Cornell, Orsay and Sasakatoon were jointly reanalysed by \emph{L.N.~Hand} et al., with the result:\cite{Ha63}
\[
r_p = 0.805(11)\,{\rm fm}\,.
\]
This value has been accepted for several decades.

\subsection{Recoil proton detection}\label{ssec:recoil}
A peculiar method was used by \emph{K.~Berkelman}, {\it et al.} at Cornell\cite{Be63}, by \emph{Frèrejaque} {\it et al.} in Orsay\cite{Fr66}, and by \emph{J.J.~Murphy} {\it et al.} in Saskatoon\cite{Mu74}: both the scattered electrons and the \emph{recoil protons} were momentum analysed and measured in coincidence making possible background-free measurements. The detection of the recoil proton at $0^\circ$ leads directly to the determination of $G_M(q^2)$, and the observation of the proton recoiling at another angle ($30^\circ$) yields $G_E(q^2)$.  At low $q$ values, the relativistic corrections to the proton are much smaller than to the electron; the collimation is easier, too.  They find \emph{rms} radii\cite{Be63, Fr66, Mu74}:
\[
r_p = 0.82(4)\,,{\rm fm}
\,,\quad r_p = 0.800(25)\,{\rm fm}
\,,\quad r_p = 0.81(3)\,{\rm fm}
\,,
\]
respectively.

\begin{figure}
\noindent\hfil\includegraphics[scale=.4]{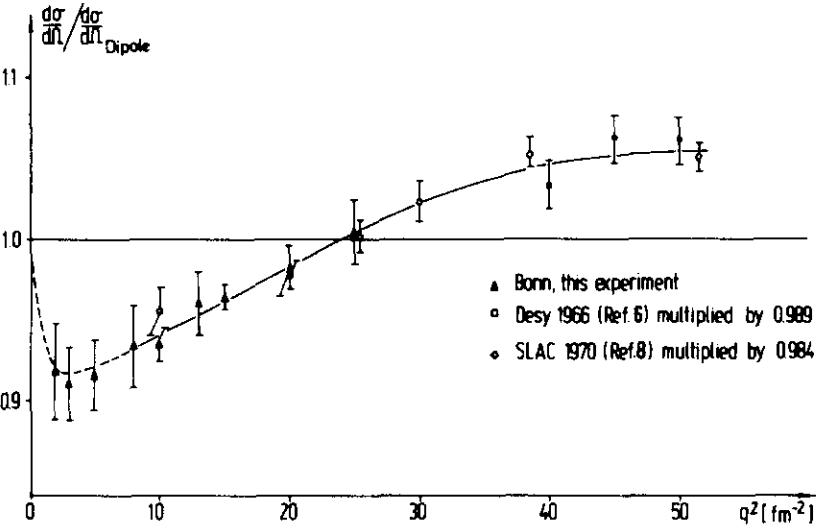}
\caption{Ratio of the experimental scattering cross section to the dipole cross-section\cite{Be71}.}
\label{fig8}
\end{figure}

The electron scattering investigations at the University of Bonn by \emph{Ch.~Berger} {\it et al.}\cite{Be68, Be71} have shown that the dipole function reproduces the trend of cross sections within about $\pm10\,\%$, see Fig.\ \ref{fig8}\cite{Be71}.

\emph{F.~Borkowski} {\it et al.} analysed their \emph{low‑$q$} measurements with different models. A fit with a sum of \emph{two double poles}
\[
G(q^2) = \sum_{i=1}^2 \frac{a_i}{(1+q^2/m_i^2)^2}\,,\quad \sum_{i=1}^2 a_i =1\,,
\]
resulted in\cite{Bo74}
\[
r_p = 0.88(3)\,{\rm fm}
\,,
\]
while the \emph{sum of four single poles}
\[
G(q^2) = \sum_{i=1}^4 \frac{a_i}{(1+q^2/m_i^2)}\,,\quad \sum_{i=1}^4 a_i =1\,,
\]
yielded\cite{Bo74}
\[
r_p = 0.92(3)\,{\rm fm}
\,.
\]
The parameters $m_i$ of two of the poles were fixed at values corresponding to the masses of the $\rho$ and $\rho'$ mesons. The other parameters have been adjusted. A dipole function would give $r_p = 0.81\,{\rm fm}$, but it failed to fit the data. In a subsequent work they extended the $q$‑region and included data from several other authors. Then the four-single-pole fit--letting $G(0)\ne 1$--, gave\cite{Bo75}
\[
r_p = 0.84(2)\,{\rm fm}
\,;
\]
demanding $G(0) = 1$ they had\cite{Bo75}:
\[
r_p = 0.88\,{\rm fm}
\,.
\]
In 1975, they reanalysed earlier data from Mainz\cite{Bo74, Bo75}, Orsay\cite{Du63} and Saskatoon\cite{Mu74}, and arrived to a joint value\cite{Bo75a}
\[
r_p = 0.87(2)\,{\rm fm}
\,.
\]

\noindent\begin{minipage}{0.05\textwidth}
  {\ }
\end{minipage}%
\begin{minipage}{0.95\textwidth}
  {\small\paragraph{ Remark:} In these years, the application of dispersion relations to the evaluation of nucleon structure measurements demanded the introduction of complex form factors and negative $q^2$ values. In order to avoid misunderstandings, in the space-like region -- for the elastic scattering -- the symbol  $Q^2$  has been often --but not generally-- applied.}
\end{minipage}
\vskip .5ex

\emph{G.~Höhler} {\it et al.} determined electromagnetic nucleon form factors from Rosenbluth plots and, independently, by fitting a dispersion ansatz to electron nucleon scattering cross sections; they derived the nucleon electric and magnetic radii\cite{Hoe76}, e.g.,
\[
r_p = 0.836(25)\,{\rm fm}
\,.
\]
For a review and analysis of Mainz data see Ref.\ \citenum{Si80a}.

In 1980, at the University of Mainz, \emph{G.G.~Simon} {\it et al.} performed electron scattering measurements on the proton in the interval $q^2 = 0.13$--$1.4\,{\rm fm}^2$  using a special hydrogen gas target\cite{Si80}. Including also results from Orsay and Saskatoon, they obtained\cite{Si80}
\[
r_p = 0.862(12)\,{\rm fm}
\,,
\]
in disagreement with the result of \emph{Hand}'s analysis. This value became a rival to that from Ref.\ \citenum{Ha63}\ (0.805\,fm), and these have been in use parallel for some time. Note that the data of Ref.\ \citenum{Si80}\ were reanalysed by \emph{Ch.W.~Wong}\cite{Wo94}, with the result:
\[
r_p = 0.877(24)\,{\rm fm}
\,.
\]

As an update and continuation of Ref.\ \citenum{Hoe76}, using the world data, \emph{P.~Mergell} {\it et al.} performed a dispersion-theoretical analysis of the nucleon electromagnetic form factors\cite{Me96}. They got a proton radius
\[
r_p = 0.847(8)\,{\rm fm}\,. 
\]
The large spread of proton \emph{rms} radius data called for a new method, independent of electron scattering.

\section{The proton radius in QED (1983--2000)}\label{sec:RpQED}
The \emph{Lamb} experiment was an important first step in the verification of QED. For testing of further, more detailed theoretical calculations, \emph{Lamb}'s radiofrequency resonance method is not suitable because the large natural width ($\sim$100\,MHz) of the \hlvl{2}{P}{} states put a limit to a more accurate measurement of the level shifts. As early as in 1983, \emph{V.G.~Pal'chikov} {\it et al.} (Moscow) measured the lifetime of the \hlvl{2}{P}{} state and the \emph{Lamb} shift in hydrogen; finally they add: \emph{-We note in conclusion that, in principle, this accuracy in the measurement of the Lamb shift ($\sim$2\,kHz) makes it possible to extract the radius of the proton within an error of  0.007\,fm from our experiment}\cite{Pa83}. Indeed, in 1985, from a new experimental $n = 2$ \emph{Lamb} shift and two different theoretical \emph{QED} calculations they obtain\cite{Pa85}
\[
r_p = 0.59(2)\,{\rm fm}
\,,\text{ and } r_p = 0.76(1)\,{\rm fm}
\,,
\]
respectively.

Applying a new idea, the \emph{separated-oscillatory-field} technique was used to obtain resonances that were significantly narrower than the natural linewidth of the transition.  With this technique \emph{S.R.~Lundeen} and \emph{F.M.~Pipkin} suceeded to improve the precision of \hlvl{2}{S}{1/2}-\hlvl{2}{P}{1/2} \emph{Lamb} shift interval in hydrogen\cite{Lu86}. They arrive at the conclusion:  \emph{-A better test of the theory will require a better value for the proton radius}.  \emph{E.W.~Hagley} and \emph{Pipkin} use the same method for the measurement of the \hlvl{2}{S}{1/2}-\hlvl{2}{P}{3/2} interval\cite{Ha94}, and conclude:  \emph{-Before an accurate comparison between theory and experiment can be made, the proton radius discrepancy must be resolved and higher order corrections to the Lamb shift must be calculated.}

With the development of quantum optics, it became possible to measure the \emph{Lamb} shift $L_1$ of the \hlvl{1}{S}{} level. The \hlvl{1}{S}{}-\hlvl{2}{S}{} transition has an unusually narrow natural linewidth of 1.3\,Hz, offering an eventual experimental resolution of 5 parts in $10^{16}$.  \emph{D.H.~McIntyre} {\it et al.}\cite{Mc89} (Stanford) were able to determine the \hlvl{1}{S}{} \emph{Lamb} shift to within 2 parts in $10^4$.  \emph{M.G.~Boshier} {\it et al.}\cite{Bo89} (Clarendon Lab., Univ.\ Oxford) improved this by a factor of  3, and declare:  \emph{-At this level of precision the discrepancy in the measurements of the proton size becomes important.}

\emph{Martin Weitz} {\it et al.} (Garching) obtain $L_1$ with an uncertainty of 1.3 parts in $10^5$, Ref.\ \citenum{We92}. The experimental data give preference to the small radius from Ref.\ \citenum{Ha63}.  In 1994, \emph{Weitz} {\it et al.} performed a precision measurement of the hydrogen and deuterium \hlvl{1}{S}{} ground state \emph{Lamb} shift\cite{We94}:  \emph{-Our experimental values differ from the theoretical predictions by $3.0\times\sigma$  for hydrogen if we base our calculations on a newer measurement of the proton charge radius ($r_p=862\,{\rm fm}$). \bf An alternative interpretation of the present measurement would be a determination of the proton and deuteron charge radii.}

\begin{figure}
\noindent\hfil\includegraphics[scale=.15]{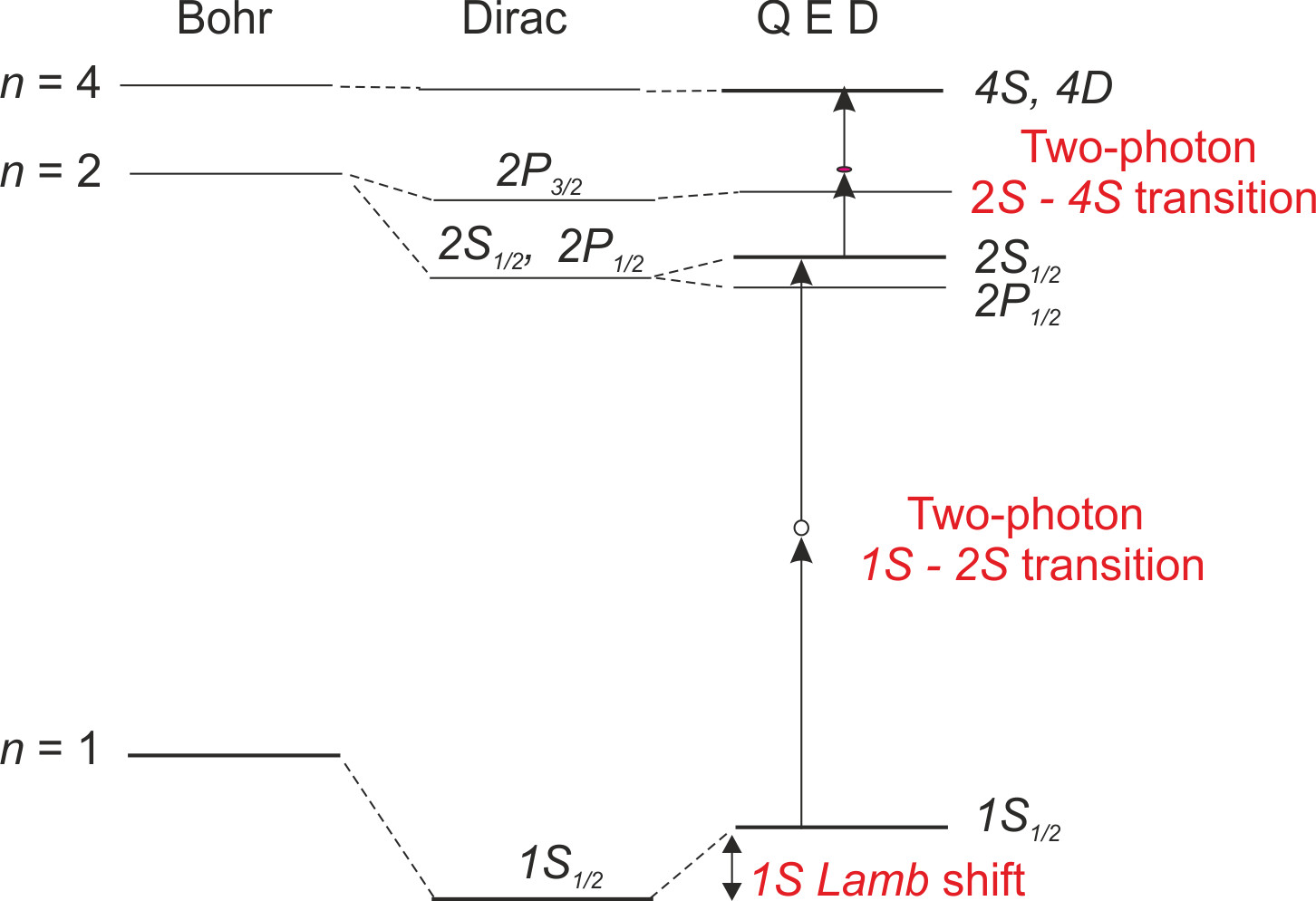}
\caption{Levels of the hydrogen atom excited by two-photon transitions\cite{We95}.}
\label{fig9}
\end{figure}

\begin{figure}
\noindent\hfil\includegraphics[scale=.15]{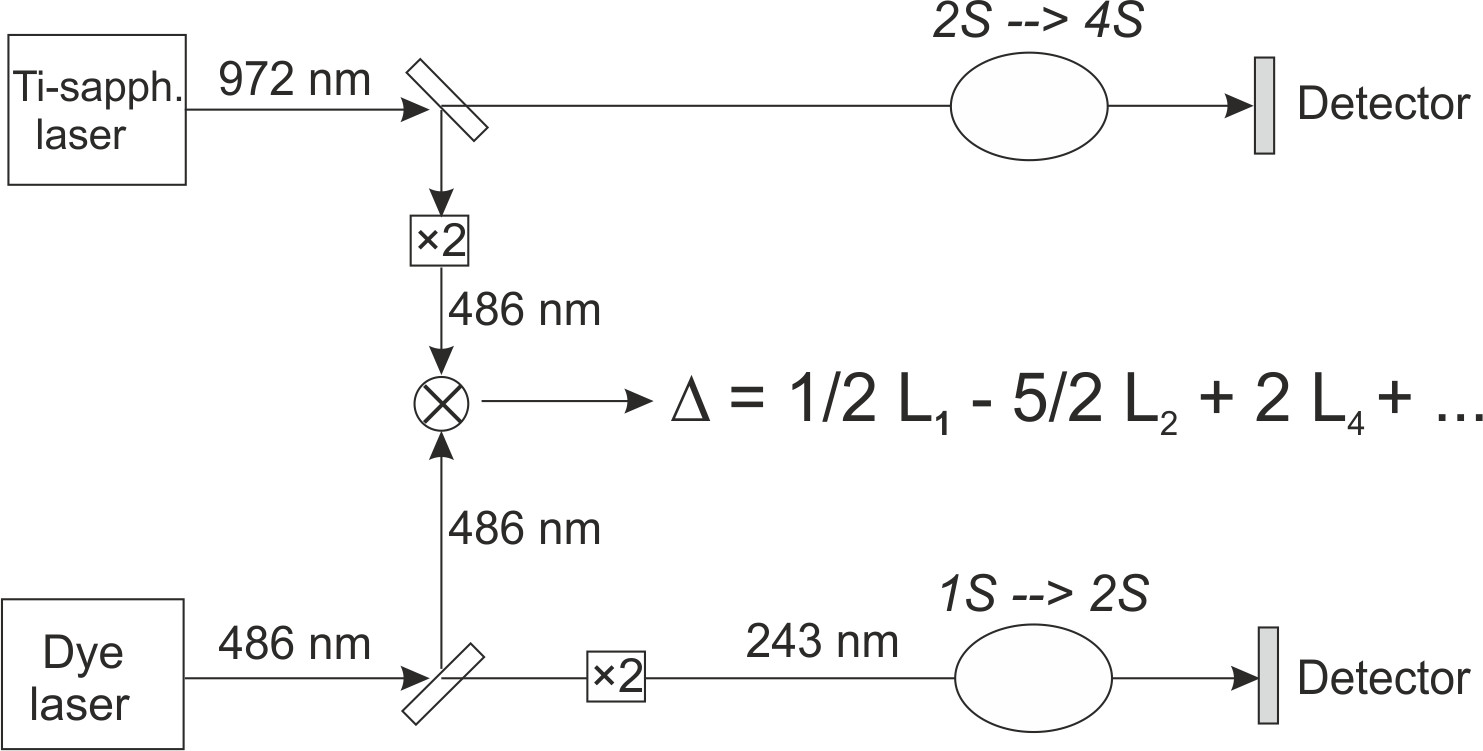}
\caption{Experimental setup for measuring the difference $\Delta$; after Ref.\ \citenum{We95}.}
\label{fig10}
\end{figure}

As the level shift depends with the inverse of the cube of the principal quantum number:  $L_n\sim 1/n^3$,  the shift $L_1$ is expected almost an order of magnitude larger than $L_2$. The measurement was performed in Garching by the group of \emph{Theodor W.~Hänsch} (Nobel Prize 2005: \emph{Passion for precision}\cite{Hae06}). The principle can easily be understood recalling the energy expression for the principal quantum number $n$:
\[
  E_n=\frac{k}{n^2}+L_n+\text{rel.\ corr.}\,.
\]
The first, \emph{Coulomb term} yields the highest contribution; $L_n$ is the Lamb shift. The small relativistic corrections can be calculated; in what follows, they will be omitted, because they do not have any role in the demonstration of the method. The differences between the energies are:
\[
  E_4-E_2=k/16+L_4-k/4-L_2+\dots\,,\quad\text{and}\quad E_2-E_1=k/4+L_2-k-L_1+\dots\,.
\]
Forming a suitable linear combination of the transitions  $\Delta n=1\to 2$ and $2\to 4$,  the Coulomb terms cancel,
\[
  \Delta = 2(E_4-E_2)-(E_2-E_1)/2 = L_1/2 -5L_2/2+2L_4\,.
\]
The shift $L_4$ is very small, an approximate theoretical estimate is sufficient to determine its value. The shift $L_2$ is an order of magnitude less than the value of $L_1$ to be determined; its value is taken from an up-to-date version of the \emph{Lamb} experiment. In this way, $L_1$ can be determined by measuring the difference $\Delta$, see Figs.\ \ref{fig9}.\ and \ref{fig10}\cite{We95}.

Using lasers, the two-photon transitions \hlvl{1}{S}{}$\to$\hlvl{2}{S}{} and \hlvl{2}{S}{}$\to$\hlvl{4}{S}{} (\hlvl{4}{D}{}) can be realized. In the lower part of the arrangement the very stable frequency (486\,nm) of a blue dye laser, after a semi-transparent mirror, is doubled to produce ultraviolet light near 243\,nm. If the \hlvl{1}{S}{}-\hlvl{2}{S}{} transition is excited, a signal is passed from the spectrometer unit to the detector. As the width of the metastable \hlvl{2}{S}{} state is very narrow, the resonance frequency can be precisely determined. In the upper part, a Ti-sapphire laser produces light near 972\,nm to excite the \hlvl{2}{S}{}-\hlvl{4}{S}{} and \hlvl{2}{S}{}-\hlvl{4}{D}{} two-photon transitions in a beam of metastable hydrogen atoms, which have been excited into the metastable \hlvl{2}{S}{} level by electron impact. The frequency of the beam reflected by the mirror is doubled and compared to the lower reflected beam in a fast photo diode. The \hlvl{1}{S}{} \emph{Lamb} shift is obtained by measuring the small frequency difference $\Delta$  as a beat node on the fast photodiode.

The experimental result for the shift $L_1$ was in agreement with the theoretically calculated value within the error limits. Table\ \ref{tab1} shows the contributions to the theoretical value. As for the uncertainty of the theory, the error of the proton charge radius is the main contribution.

\begin{table}
  \tbl{Contributions to the theoretical \emph{QED} value of the \hlvl{1}{S}{} \emph{Lamb} shift\cite{We95}.}{
    \begin{tabular}{|l|r|}
      \hline
      {\bf Energy contributions} &  {\bf MHz}\\
      \hline\hline
      Self energy                & 8396.456(1)\\
      Vacuum polarization        & ‑215.168(1)\\
      Higher order QED           &  0.724{\bf (24)}\\
      Radiative recoil corr.     & ‑12.778(6)\\
      Non-radiative recoil corr. & 2.402(1) \\
      Proton charge radius       & 1.167{\bf (32)}\\
      {\bf Sum of theoretical contributions} & {\bf 8172.802(40)}\\
      \hline
      \emph{\bf Experimental value} & \emph{\bf 8172.874(60)}\\
      \hline
    \end{tabular}
    }
  \label{tab1}
\end{table}

\emph{Th.~Udem} {\it et al.} measured the absolute frequency of the hydrogen \hlvl{1}{S}{}-\hlvl{2}{S}{} transition improving the previous accuracy by almost two orders of magnitude, and derived the \hlvl{1}{S}{} \emph{Lamb} shift\cite{Ud97}. The agreement with the theory was only moderate with any of the proton radii from the literature. Their conclusion is  \emph{-Our experiment can be interpreted as a measurement of the proton rms charge radius, yielding}\cite{Ud97}
\[
  r_p=0.890(14)\,{\rm fm}
\,,
\]
\emph{provided that the theoretical calculations are correct.}

In Paris, \emph{C.~Schwob} {\it et al.} from the precise measurement of the \hlvl{2}{S}{}-\hlvl{12}{D}{} transition deduced\cite{Sc99a}
\[
  r_p=0.900(16)\,{\rm fm}
\,.
\]

In 2000, a further theoretical contribution to hydrogen energy levels was calculated by \emph{Kirill Melnikov} (Stanford) and \emph{Timo van Ritbergen} (Karlsruhe);  applying this to \emph{Schwob}'s experimental data, yielded\cite{Me00}
\[
  r_p=0.883(14)\,{\rm fm}
\,.
\]

It should be stressed that these derivations use the energies of \emph{bound} electron states, differently of \emph{free electron scattering}. Undiscovered systematic errors of the electron scattering experiments do not have any effect on the \emph{Lamb} shift measured in the $\Delta$ experiment.

\begin{figure}
\noindent\hfil\includegraphics[scale=.1]{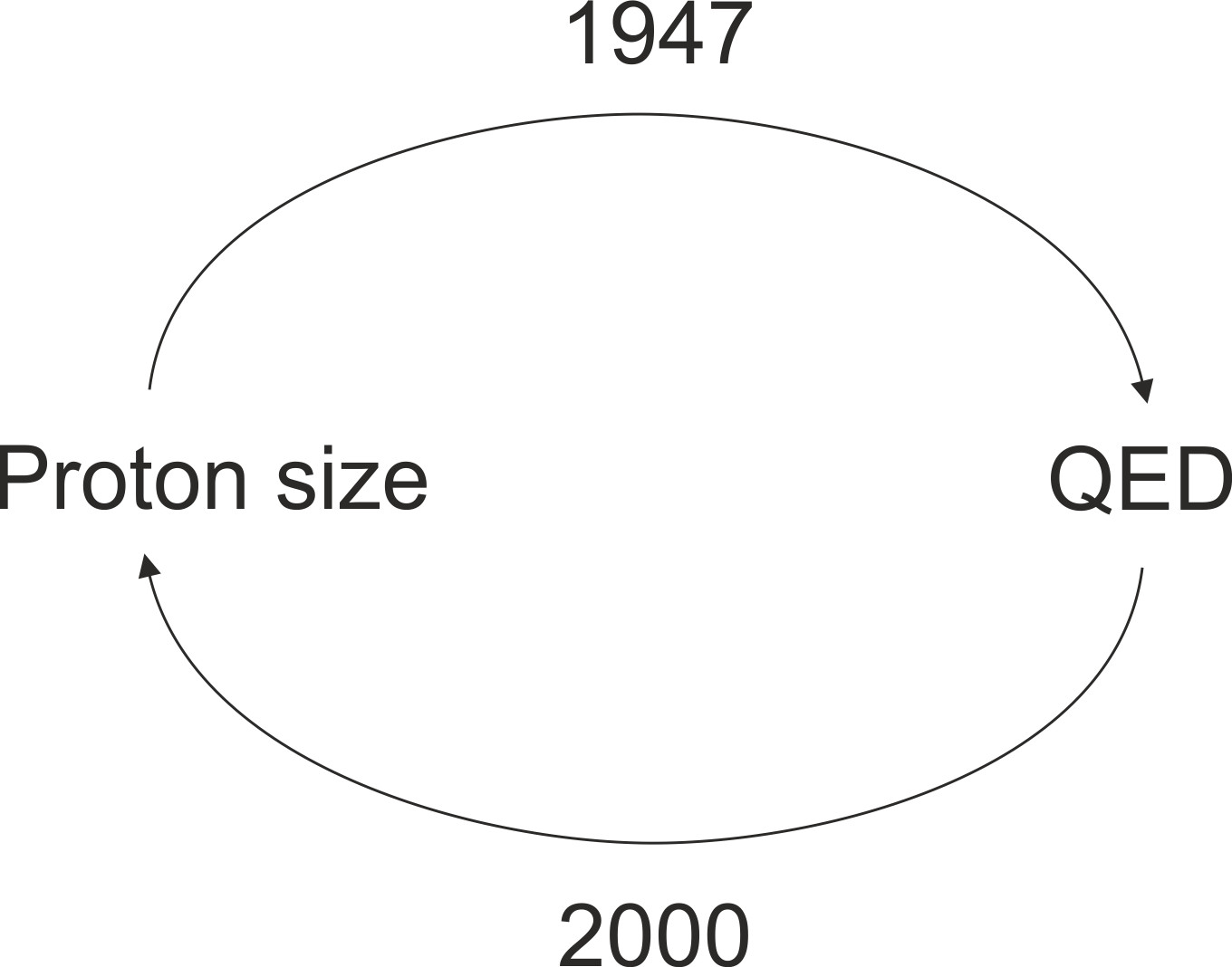}
\caption{Historical route of the search for the proton size.}
\label{fig11}
\end{figure}

We can draw our route on Fig.\ \ref{fig11}:  \emph{Lamb} searched for the size of the positive meson cloud, i.e., for the radius of the proton. He discovered $L_2$, the effect of the electromagnetic radiation field on the \hlvl{2}{S}{} level in the hydrogen atom, i.e., the shift supporting the idea of quantum electrodynamics. On the other hand, \emph{Hänsch} with his group investigated the shift $L_1$;  from their experiment, --as a valuable by-product--,  the proton charge radius emerged.

This mutual assistance rings well but for the modern test of \emph{QED} an independent proton size is necessary. Otherwise, the situation is similar to the tale from \emph{Baron Münchausen} (born 1720, 300 years ago) who pulled himself and his horse out of the marsh by seizing his own hair.

\section{If QED then some island; --also reanalyses (2000--2010)}\label{sec:island}
As in 1947 on Shelter Island, half a century later on Isola di San Servolo (Venice), a few physicists met to discuss ever more stringent tests of quantum electrodynamics\cite{Be14}. A necessary ingredient to this test was a more precise proton size. Here, two young physicists, \emph{Jan C.~Bernauer} and \emph{Randolph Pohl} decided to tackle the problem in two different ways. The work needed a decade, and the result was the \emph{proton radius puzzle}.

In the meantime, however, several reanalyses of earlier electron scattering data took place.

In 2003, \emph{Ingo Sick} reanalysed the experimental data for $q^2=0\dots16\,{\rm fm}^{-2}$ {\ }\cite{Si03}. He took into account the Coulomb distortion neglected earlier, and wrote the electric form factor in the \emph{continued fraction} form:
\[
  G_E(q^2) = \frac{1}{1+\displaystyle\frac{b_1q^2}{1+\frac{b_2q^2}{1+\frac{\dots}{\dots}}}}\,.
\]
The parameters $b_1$ and $b_2$  are determined by a fit to experimental data. These parameters are directly connected to the coefficients of the power series in $q^2$:
\[
  b_1 = \frac{\langle r^2\rangle}{3!}\,,\quad\text{and}\quad b_1^2+b_1b_2 = \frac{\langle r^4\rangle}{5!}\,.
\]
Applying the above procedure, he got a radius value\cite{Si03}
\[
  r_p=0.895(18)\,{\rm fm}
\,.
\]

\emph{Peter G.~Blunden} and \emph{Ingo Sick} applied a two-photon correction to the above data set, yielding\cite{Bl05}
\[
  r_p=0.897(18)\,{\rm fm}
\,.
\]

\emph{M.A.~Belushkin} {\it et al.} applied dispersion analysis to nucleon form factors including meson continua using two approaches: explicit \emph{pQCD} and superconvergence (\emph{SC}).  Nucleon radii were also extracted; the results are for the proton\cite{Be07}:
\[
  r_p=0.830(^{+5}{}_{-8})\,{\rm fm}
\,,\quad\text{and}\quad  r_p=0.844(^{+8}{}_{-4})\,{\rm fm}
\,,
\]
for the two approaches, respectively.

\emph{D.~Borisyuk} also performed a reanalysis of electron scattering data, which --as stated--, was more accurate than that of \emph{Sick}.  He obtained the result\cite{Bo10}
\[
  r_p=0.912(9)_{\rm stat}(7)_{\rm syst}\,{\rm fm}
\,.
\]

In 2010, \emph{Richard J.~Hill} and \emph{Gil Paz} publish a model-independent extraction of the proton charge radius combining constraints from analyticity with experimental electron-proton scattering data; the result is\cite{Hi10}:
\[
  r_p=0.870(23)(12)\,{\rm fm}
\,.
\]
Using also electron-neutron scattering data, they have the value\cite{Hi10}
\[
  r_p=0.880(^{+17}{}_{-20})(7)\,{\rm fm}
\,,
\]
while adding $\pi\pi\to NN$ data\cite{Hi10}:
\[
  r_p=0.871(9)(2)(2)\,{\rm fm}
\,.
\]

\section{A puzzle and the efforts to resolve it (2010--2020)}\label{sec:puzzle}
\begin{figure}
  \hfil\begin{tabular}{cc}
         \includegraphics[scale=.5]{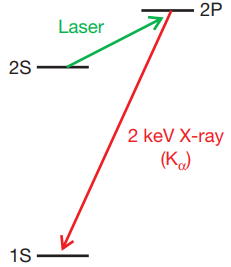} & \includegraphics[scale=.3]{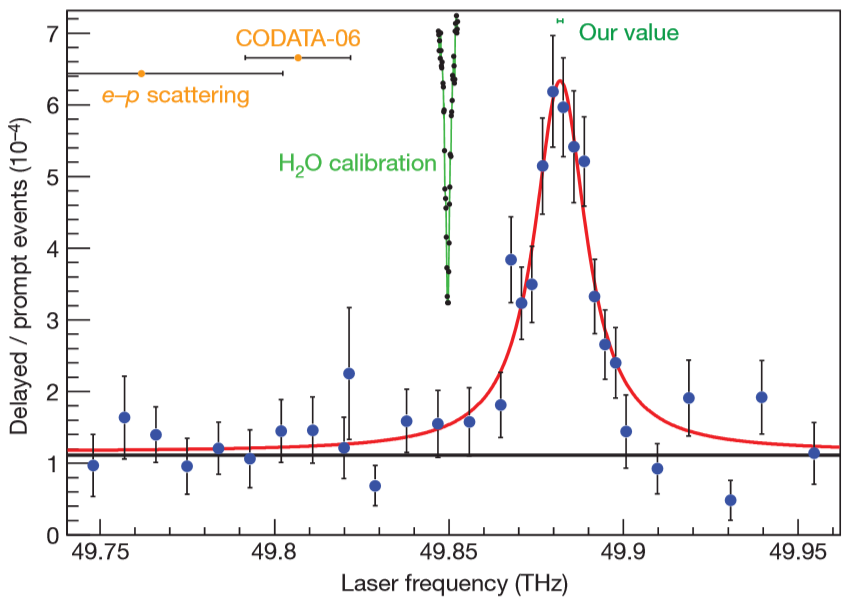} \\
         {\small (a) } & {\small (b)\cite{Po10}}\\
       \end{tabular}
       \caption{}\label{figNN}
\end{figure}

Preparations were done by an international group in PSI (Paul Scherrer Institut, Villigen) for the determination of the proton radius by measuring the \hlvl{2}{S}{} \emph{Lamb} shift in \emph{muonic} hydrogen\cite{Ta99}. The stopping of negative muons in a small volume of hydrogen gas at low densities has been accomplished in two experiments. The existence of long-lived $\mu p$(\hlvl{2}{S}{}) atoms made it possible to use an external laser to excite it into the \hlvl{2}{P}{} state (Fig.\ \ref{figNN}a), and measure the laser frequency where this transition takes place (Fig.\ \ref{figNN}b). The transition energy contains the proton \emph{rms} charge radius.

It was expected that this experiment will be a verification of the \emph{rms} radii values obtained from electron-scattering and electron-spectroscopy;  no ``scandal'' was plotted.

\begin{figure}
\noindent\hfil\includegraphics[scale=.4]{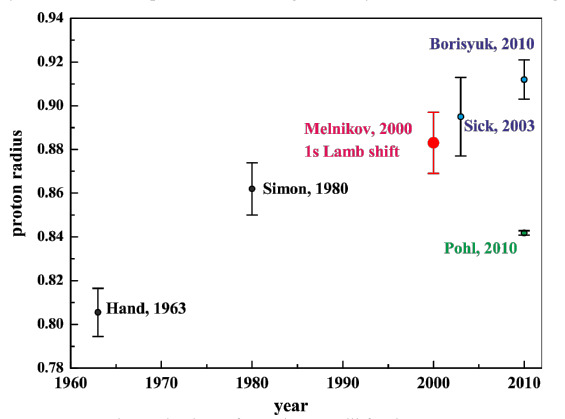}
\caption{Some experimental values of rms charge radii for the proton up to 2010\cite{An13}. {\bf Black}: electron scattering, exp.; {\color{red}\bf Red}: electronic $L_1$ Lamb shift;
{\color{blue}\bf Blue}: reanalyses of electron scattering.  {\color{green}\bf Green}: muonic $L_2$ Lamb shift.}
\label{fig12}
\end{figure}

And then came the bolt from the blue! They obtained a value\cite{Po10}
\[
  r_p=0.84184(67)\,{\rm fm}
\,,
\]
more precise than any of the former experiments, but a significantly different value; see Fig.\ \ref{fig12}.

This surprising development gave rise to active interest and attempts to interpretation. If the deviation is not caused by some undiscovered systematic experimental error or calculation mistake then it would be reasonable to assume that the muon is not a simple ``fat electron''. This would have far-reaching consequences: the validity of the Standard Model of particle physics would be questionable, see, e.g., Ref.\ \citenum{Fl10}. The atmosphere may be characterized by listing the titles of some papers published within one year:
\begin{itemize}
\item \emph{``Quantum electrodynamics: A chink in the armour?''}\cite{Fl10}.
\item \emph{``QED is not endangered by the proton’s size''}\cite{DR10}.
\item \emph{``QED confronts the radius of the proton''}\cite{DR11}.
\item \emph{``The RMS charge radius of the proton and Zemach moments''}\cite{Di11}.
\item \emph{``Lamb shift in muonic hydrogen-I. Verification and update of theoretical predictions''}\cite{Je11}.
\item \emph{``Lamb shift in muonic hydrogen-II Analysis of the discrepancy of theory and experiment''}\cite{Je11a}.
\item \emph{``Natural Resolution of the Proton Size Puzzle''}\cite{Mi11}.
\item \emph{``Higher-order proton structure corrections to the Lamb shift in muonic hydrogen''}\cite{Ca11}.
\item \emph{``Troubles with the Proton rms-Radius''}\cite{Si11}.
\end{itemize}
The list may not be complete.

\emph{J.C.~Bernauer} {\it et al.} (A1 Collaboration, Mainz) published new precise measurement of the elastic electron-proton scattering cross section covering $Q^2$ from 0.004 to 1 $({\rm GeV}/c)^2$ {\ }\cite{Be10, Be14a}.  More than 1400 cross sections were measured with statistical errors below 0.2\,\%. The spectrometer angles were varied only in small steps so that the same scattering angle is measured up to four times with different regions of the spectrometer acceptance, and parts of the angular range were measured with two spectrometers.

The analysis differs in some points from the customary. The calibration problem present in any determination of absolute cross sections has been overcome by an explicit luminosity measurement using an extra spectrometer. This makes a precise determination of the absolute normalization for all measurements possible.

Only the objectively proven radiative corrections have been applied. However, an empirical form has been derived from the inconsistency of the $G_E$ and $G_M$ data, extracted from measurements with polarized and unpolarized electrons, respectively, which may be interpreted by radiative corrections as TPE or other physics. The new method has been applied to the world data set together with the data with this paper. The charge radius of the proton so determined is\cite{Be10}
\[
  r_p=0.879(5)_{\rm stat}(4)_{\rm syst}(2)_{\rm model}(4)_{\rm group}\,{\rm fm}
\,.
\]
In the answer to the Comment from \emph{J.~Arrington}\cite{Ar11}, this value was modified to\cite{Be11}
\[
  r_p=0.876(8)\,{\rm fm}
\,.
\]
The authors conclude:\\
\emph{\bf ‑Despite all these efforts, we do not see a way to reconcile our result with those from muonic hydrogen.}

\emph{I.T.~Lorentz} {\it et al.} (Univ. Bonn, Jülich) analysed \emph{Bernauer}'s electron scattering data set using a dispersive framework that respects the constraints from analyticity and unitarity on the nucleon structure, and found a \emph{small} proton radius\cite{Lo12}:
\[
  r_p=0.84(1)\,{\rm fm}
\,.
\]

\subsection{Polarization transfer from polarized electron to recoil proton}\label{ssec:PolarizationTransfer}
\emph{G.~Ron} {\it et al.} (Jefferson Collab.) performed a \emph{polarization transfer} measurement with longitudinally polarized electrons hitting on protons to determine the proton form factor ratio $\mu_pG_E/G_M$ {\ }\cite{Ro11}. The polarization measurements with a focal plane polarimeter allow for a much better separation of  $G_E$  and  $G_M$.  In addition, the contribution of two-photon exchange effects, which have a large impact on the extractions from the unpolarized cross-section measurements, have less impact on the polarization measurements.  The derived rms charge radius of the proton is\cite{Ro11}
\[
  r_p=0.875(10)\,{\rm fm}
\,.
\]
Combined with the cross-section measurements from Mainz\cite{Be10}, results in\cite{Ro11}
\[
  r_p=0.8772(46)\,{\rm fm}
\,.
\]

\emph{X.~Zhan} {\it et al.} (Jefferson Collab.) performed a high-precision measurement of the proton elastic form factor ratio  $\mu_p G_E/G_M$  at low $Q^2$ using recoil polarimetry. They obtained the proton electric radius\cite{Zh11}:
\[
  r_p=0.875(10)\,{\rm fm}
\,.
\]

\emph{Ingo Sick} calls the attention to the fact that the value of the proton \emph{rms} radius determined from electron scattering data depends strongly on the density $\rho(r)$ at large radii $r$ {\ }\cite{Si11}. This density is poorly constrained by scattering data. Supplementing the $(e,e')$ data with the large-$r$ shape of $\rho(r)$ resulting from the \emph{Fock} components $(n+\pi,\dots)$, which dominate the large-$r$ behaviour, produces a more reliable value for the radius\cite{Si12}:
\[
  r_p=0.886(8)\,{\rm fm}
\,.
\]

\emph{Aldo Antognini} {\it et al.} measured the $2{\rm S}_{1/2}^{F=0}-2{\rm P}_{3/2}^{F=1}$ transition frequency and reevaluated the $2{\rm S}_{1/2}^{F=1}-2{\rm P}_{3/2}^{F=2}$  transition frequency.  From the measurements, they extracted the proton charge radius\cite{An13a}:
\[
  r_p=0.84087(39)\,{\rm fm}
\,,
\]
close to the earlier result\cite{Po10}.

In 2013, a fundamental paper is published reviewing the theory of the \hlvl{2}{S}{}-\hlvl{2}{P}{} Lamb shift and \hlvl{2}{S}{} hyperfine splitting in muonic hydrogen combining the published contributions and theoretical approaches. The \emph{theoretical prediction} for the \emph{Lamb} shift in muonic hydrogen, Eq.\ (35) in Ref.\ \citenum{An13b}:
\[
  \Delta E_L= 206.0668(25)-5.2275(10)\times r_p^2\,{\rm meV}\,.
\]
The theoretical prediction of these quantities is necessary for the determination of both the proton charge and the Zemach radii from the two \emph{2}{S}{}-\emph{2}{P} transition frequencies measured in muonic hydrogen.

\begin{figure}
\noindent\hfil\includegraphics[scale=.45]{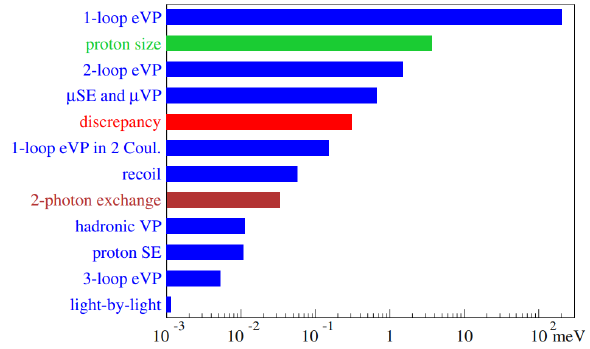}
\caption{Contributions to the muonic \emph{Lamb} shift\cite{Po14}. Note the logarithmic scale. The proton size is the second largest term. The TPE has recently gained considerable interest.}
\label{fig13}
\end{figure}

The exciting atmosphere of these years is excellently reflected in the review paper of \emph{Randolf Pohl} {\it et al.}\cite{Po13}. It is like a contemporary document: practically every way to the solution is open, see the \emph{Conclusions} on page 45.; worth to read even today.  This is followed by another paper\cite{Po14} containing an instructive figure demonstrating the theoretical contributions to the \emph{Lamb} shift in $\mu p$, Fig.\ \ref{fig13}. Note the logarithmic scale. The theory has been carefully checked by various authors, but no large missing or wrong \emph{QED} term has been found.

After almost one and a half decade, the two young men who met on Isola di San Servolo, sum up the situation\cite{Be14}:
\begin{quote}
  \emph{``Four years after the puzzle came to life, physicists have exhausted the straightforward explanations. We have begun to dream of more exciting possibilities.}
  
  \emph{Do we really understand how the proton reacts when the muon pulls on it? The electrostatic force of the muon deforms the proton, … The crooked proton slightly alters the \hlvl{2}{S}{} state in muonic hydrogen. Most people think that we understand this effect, but the proton is such a complicated system that we may have missed something.}
  
  \emph{The most exciting possibility is that these measurements might be a sign of new physics that go beyond the so-called Standard Model of particle physics.}
  
  \emph{On the other hand, physicists already have another muon puzzle to solve. … Tellingly, the muon's magnetic moment does not match the QED calculations. Perhaps new physical phenomena will explain both the proton radius measurement and the muon's anomalous magnetic moment.''}\cite{Be14}.
\end{quote}

New measurements at MAMI\cite{Be14a} are the continuation of the earlier\cite{Be10}, but extend the momentum transfer interval down to $0.003\,{\rm GeV}^2$, applying a new determination of the electric and magnetic form factors through a new fit of these data, together with the previous world data including the results from polarization measurements. The earlier result\cite{Be14a}
\[
  r_p=0.879(5)_{\rm stat}(4)_{\rm syst}(2)_{\rm model}(4)_{\rm group}\,{\rm fm}
\,,
\]
is confirmed.

\begin{center}
  {\bf There seems no way to reconcile these results with those from muonic hydrogen.}
\end{center}

Are there undiscovered systematic errors associated with proton radius extractions? \emph{John Arrington} discusses sources of systematic uncertainty and model dependence in the radius extractions\cite{Ar15}. He finds several issues that larger uncertainties than previously quoted may be appropriate, but does not find any corrections which would resolve the proton radius puzzle.

As a continuation, in the same Proceedings \emph{John Arrington} and \emph{Ingo Sick} re-examine the charge radius extractions from electron scattering measurements. They provide a recommended value\cite{Ar15a}
\[
  r_p=0.879(11)\,{\rm fm}
\,,
\]
based on a global examination of data. The uncertainties include contributions to account for tensions between different data sets and inconsistencies between radii using different extraction procedures. Special attention is payed to the two-photon exchange corrections.

\emph{J.J.~Krauth} {\it et al.} extended the search to muonic atom and hydrogen spectroscopy\cite{Kr17}. The questions and the answers in short:
\begin{quote}
  \emph{Correctness of muonic hydrogen experiment?}
  
  The rate of 6\,events/h observed at resonance made the search time consuming. Eventually, the statistical uncertainty limited the total experimental accuracy making the results less prone to systematic errors.
  
  \emph{Correctness of hydrogen spectroscopy?}
  
  The $r_p$ value extracted by pairing the \hlvl{1}{S}{}-\hlvl{2}{S}{} and the \hlvl{2}{S}-\hlvl{8}{D}{} transitions is showing a $3\sigma$  deviation from $\mu p$ while all the others differ only by  $\le 2\sigma$.  A $4\sigma$  discrepancy between $r_p$ from $\mu p$ and H spectroscopy alone emerges only after averaging all measurements in H.

  \emph{Correctness of the muonic hydrogen theory?}

  To extract $r_p$ from the measurement of $\mu p$ the following theoretical prediction was used:
  \[\begin{aligned}
    E_{\mu p} (2{\rm S}-2{\rm P})=&206.0336(15)\,{\rm meV}-5.2275(10)\,{\rm meV}/{\rm fm}^2\times r_p^2\\&+0.0332(20)\,{\rm meV}\,,
  \end{aligned}\]
  the last term standing for the two-photon exchange. Both the purely bound-state \emph{QED} part and the \emph{TPE} contribution are sound and have been confirmed by various groups.

  \emph{Correctness of electron proton scattering?}
  
  Because the form factor $G_E$ can be measured only down to a minimal $Q^2$, a fit with an extrapolation to $Q^2=0$ is needed to deduce $r_p$. Fit functions given by truncated general series expansions have been used, some authors additionally enforcing analyticity and coefficients with perturbative scaling, some others constraining the low-$Q^2$ behaviour of the form factor, or modeled the large-$r$ behaviour by the least–bond \emph{Fock} component of the proton. The more traditional analyses obtain $r_p$ values systematically larger than obtained by other authors that restricted their fits to very low $Q^2$ and used low-order power series.  The conclusion is:

  \emph{\bf -The $r_p$ extraction from electron scattering remains a controversial subject.}
\end{quote}

\section{Puzzle resolved? (2010--2020)}\label{sec:PuzzleResolved}

An interesting approach is that of \emph{Marko Horbatsch} {\it et al.} (York Univ., Toronto)\cite{Ho17}. They determine $r_p$ from a fit to low-$Q^2$ electron-proton scattering cross-section data from MAMI, with the higher moments fixed (within uncertainties) to the values predicted by chiral perturbation theory. They obtain\cite{Ho17}
\[
  r_p=0.855(11)\,{\rm fm}
\,.
\]

\emph{Axel Beyer} {\it et al.} (Garching) measured the \hlvl{2}{S}{}-\hlvl{2}{P}{} transition frequency in hydrogen\cite{Be71}. They used a well-controlled cryogenic source of 5.8\,K cold \hlvl{2}{S}{} atoms. Here, Doppler-free two-photon excitation is used to almost exclusively populate the  $2{\rm S}_{1/2}^{F=0}$ Zeeman sublevel without imparting additional momentum on the atoms. With the \hlvl{1}{S}{}-\hlvl{2}{S}{} transition frequency measured earlier, the Rydberg constant $R_\infty$ and $r_p$ are derived, simultaneously.  These two quantities are in very strong correlation, with a correlation coefficient  0.9891.  For the proton \emph{rms} charge radius they obtained\cite{Be17}
\[
  r_p=0.8335(95)\,{\rm fm}
\,.
\]
This value is in good agreement with the muonic atom value, but in discrepancy of  $3.3\times\sigma$ to the H spectroscopy world data (Ref.\ \citenum{Mo16}, Table~XXIX.).

\emph{Hélène Fleurbaey} {\it et al.} (Paris) measured the \hlvl{1}{S}{}-\hlvl{3}{S}{} transition frequency in hydrogen; combining it with the \hlvl{1}{S}{}-\hlvl{2}{S}{} transition frequency, the result is\cite{Fl18}
\[
  r_p=0.877(13)\,{\rm fm}
\,.
\]
This work is being contested by fluorescence-detection work in Garching, which is achieving higher precision\cite{Ho20}.

\emph{J.M.~Alarcón} {\it et al.} present a novel predictive theoretical framework, for the extraction of proton radius from elastic form factor (FF) data, implementing analyticity: dispersively improved chiral effective field theory (DI$\chi$EFT)\cite{Al19}. They express the spacelike proton FF predicted by the theory in a form such that it contains the radius as a free parameter, and obtain\cite{Al19}
\[
  r_p=0.844(7)\,{\rm fm}
\,.
\]

An ingenious method, the frequency-offset separated oscillatory field (FOSOF) technique was used by \emph{N.~Bezginov} {\it et al.} (York Univ. Toronto)\cite{Be19} to measure the energy difference between the \hlvl{2}{S}{1/2}($F=0$) and \hlvl{2}{P}{1/2}($F=1$) states in hydrogen. From the measurement a value of the proton \emph{rms} charge radius can be deduced\cite{Be19}:
\[
  r_p=0.833(10)\,{\rm fm}
\,.
\]

\begin{figure}
\noindent\hfil\includegraphics[scale=.45]{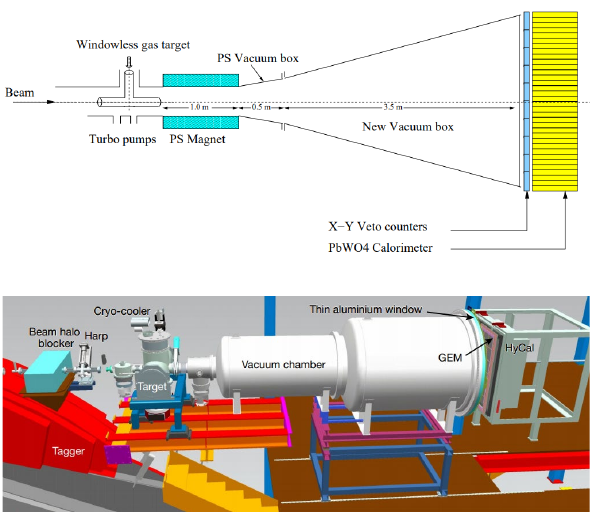}
\caption{The PRad experimental arrangement (Hall B Jefferson Lab)\cite{Kh12, Xi19}.}
\label{fig14}
\end{figure}

\emph{W.~Xiong} {\it et al.} (PRad Collab. Jefferson Lab.) implemented three major improvements over previous e‑p experiments, see Fig.\ \ref{fig14}.\cite{Xi19}. First, the large angular acceptance ($0.7^\circ$-$7.0^\circ$) of the hybrid calorimeter (HyCal) consisting of 580 Pb-glass and 1200 PbWO\textsubscript{4} crystals, a plane ($1.16\times 1.16\,{\rm m}^2$) made of two high-resolution X-Y gas electron multiplier (GEM) with $72\,\mu{\rm m}$ resolution, enabled large $Q^2$ coverage spanning two orders of magnitude in the low-$Q^2$ range. The fixed location eliminated the many normalization parameters that plague magnetic-spectrometer-based experiments in which the spectrometer must be physically moved to many different angles.

Second, the extracted $e-p$ cross-sections were normalized to the well-known QED process $e^-e^-\to e^-e^-$ (\emph{Møller} scattering from atomic electrons), which was measured simultaneously alongside $e-p$ scattering, using the same detector acceptance. This led to substantial reduction in the systematic uncertainties of measuring the $e-p$ cross sections. 

Third, using a cryo-cooled hydrogen gas flow target, the background generated from the target windows, one of the dominant sources of systematic uncertainty in all previous $e-p$ experiments, was highly suppressed.

To find the slope of $G_E(Q^2)$ at $Q^2=0$, various functional forms were examined for their robustness; finally, the multi-parameter \emph{Rational}(1,1) function was applied with the result\cite{Xi19}:
\[
  r_p=0.831(7)_{\rm stat}(12)_{syst}\,{\rm fm}
\,.
\]

\emph{Timothy B.~Hayward} and \emph{Keith A.~Griffoen} (College of William and Mary, Williamsburg) examined several low-$Q^2$ elastic $ep$ and $ed$ scattering data sets using various models to extract the proton and deuteron \emph{rms} radii\cite{Ha20}. They demonstrate how the discrepancy between small and large radii arises in fits to data.  Depending on how the linear/quadratic ambiguity is resolved, reasonable fits can yield radii from 0.84 to 0.88\,fm, the smaller result being more likely.

\section{Waiting for the Kiss of the Muse (2020--$\dots$)}\label{sec:KissOfTheMuse}
The proton radius puzzle is being addressed by new experimental efforts.

\subsection{MUSE}\label{ssec:MUSE}
\emph{J.~Arrington} {\it et al.} proposed to measure $e^\pm p$, $\mu^\pm p$ and $\pi^\pm p$ scattering, which will allow a determination of the consistency of the $\mu p$ interaction with the $ep$ interaction\cite{Ar12}. The {\bf MU}on proton {\bf S}cattering {\bf E}xperiment (MUSE) at the PSI $\pi$M1 beam line is an effort to expand the comparisons by determining the proton radius through muon scattering, with simultaneous electron scattering measurements. If the $\mu p$ scattering is consistent with muonic hydrogen measurements but inconsistent with $ep$ scattering measurements, this would provide strong evidence for beyond standard model physics. As presented by \emph{R.~Gilman} {\it et al.} (The MUSE Collaboration)\cite{Gi17}:
\begin{quote}
  \emph{Our intent is}
  \begin{itemize}
  \item \emph{to directly compare $ep$ to $\mu p$ at the sub-percent level, in simultaneous measurements, more precisely than done before, and at much lower $Q^2$}
  \item \emph{to compare scattering of positive vs.\ negative charged particles to test two-photon exchange effects in both reactions at the sub-percent level, more precisely than done before,}
  \item \emph{to extract radii from both reactions, the first significant $\mu p$ scattering radius determination, at roughly the same level as done in previous scattering experiments.}
  \end{itemize}
\end{quote}

\subsection{FAMU}\label{ssec:FAMU}
The FAMU ({\bf F}isica degli {\bf A}tomi {\bf MU}onici) experiment projected by \emph{C.~Pizzolotto} {\it et al.} (INFN, RIKEN-RAL) aims to measure the \emph{hyperfine splitting of the muonic hydrogen ground state}\cite{Pi20}. From this experiment the proton Zemach radius can be derived and this will shed light on the determination of the \emph{proton charge radius}. The FAMU experiment takes place at the RIKEN-RAL muon beam facility at the Rutherford-Appleton Laboratory, in the Oxfordshire (UK). Since 2013, the FAMU team had four very fruitful data taking sessions at the facility demonstrating that the proposed method is suitable for the proposed task.

\subsection{ALPHA}\label{ssec:ALPHA}

The first results of the \emph{ALPHA Collaboration}\cite{AL20} represent an important step towards precision measurements of the fine structure and the \emph{Lamb} shift in the \emph{antihydrogen spectrum} as tests of the charge-parity-time symmetry and towards the determination of the \emph{antiproton charge radius}.

\subsection{COMPASS++/AMBER\cite{Be20}}\label{ssec:COMPASS}
They will employ a hydrogen Time-Projection-Chamber (TPC) to measure the muon-proton cross section in the $Q^2$ range of 0.001 to 0.0037\,(GeV/$c^2$). The experiment aims to measure both outgoing lepton as well as the recoiling proton, and will use both muon charges.

\subsection{MAINZ\cite{Be20}}\label{ssec:MAINZ}
\paragraph{a.} In the A2 hall, an experiment will make use of a hydrogen TPC to measure the recoil protons for a $Q^2$ range from 0.001 to 0.04\,(GeV/$c^2$). This technique will have different systematic sensivitiy than the more usual detection of thescattered electrons. The experiment aims for 0.2\% absolute and 0.1\% relative errors.

\paragraph{b.} In a combined experiment at A1@MAMI and the MAGIX@MESA, currently under construction, an updated version of the original Mainz measurement will be performed. The main improvement is the new target system, which will exchange the cryogenic cell with a hydrogen cluster-jet target.

\subsection{ULQ2\cite{Be20}}
This project at Tohoku University, Sendai, Japan, aims to measure the electron-proton cross section in the $Q^2$ range of 0.0003 to 0.008\,(GeV/$c^2$) using beam energies between 20 and 60\,MeV. The experimenters plan to use a CH\textsubscript{2} target to achieve an absolute measurement on the 3~per mil level, by measuring relative to the well known carbon cross section.

\section{The glory of ``recent results''}\label{sec:TheGlory}
\begin{figure}
\noindent\hfil\includegraphics[scale=.7]{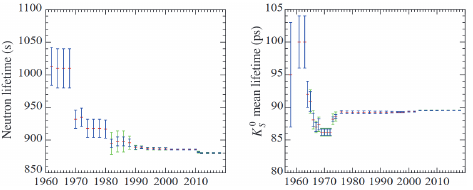}
\caption{History of measured particle data\cite{PDG18}.}
\label{fig15}
\end{figure}

Do not forget that once each of the experiments produced the \emph{``most recent''} result.  In all probability, the authors of the experiments proceeded according to their best knowledge: \emph{arte legis}.  Still, in different laboratories, different groups of physicists obtained significantly different values.  This is rather rule than exception.  Therefore, the result from muonic \hlvl{2}{S}{}-\hlvl{2}{P}{} \emph{Lamb} shift with its imposing precision should also be confirmed by \emph{independent} experiments in different laboratories by different groups.
\begin{quote}
  \emph{-Then, which  is the final result?}
  
  \emph{-Ask history:} Fig.\ \ref{fig15}.
  
  \emph{-``You Never Can Tell''} (G.B.~Shaw)
\end{quote}

\paragraph{Acknowledgements} The author thanks \emph{Mrs.~Dóra Zolnai} for technical support.
\paragraph{Personal note} This paper is dedicated to the memory of \emph{Vladimir Naumovich {\bf Gribov}}. I did not know him personally, but there are some chance coincidences in our lives. We were born in the same year. He arrived at Gatchina in 1971; the same year I felt the active scientific atmosphere of that Institute. He got to Moscow in 1980. Maybe we passed by each other at the Red Square, Tretyakov Gallery or GUM, me as a tourist and he as a local passer-by. With this--there is the end of coincidences: he was a brilliant theorist $\dots$.

\end{document}